\documentclass[12pt]{iopart} % for single column layout
\usepackage[utf8]{inputenc}
\usepackage[english]{babel}
\usepackage{footmisc}
\expandafter\let\csname equation*\endcsname\relax
\expandafter\let\csname endequation*\endcsname\relax
\usepackage{amsmath, amssymb}
\usepackage{cases}
\usepackage{graphicx}
\usepackage{upgreek} 
\usepackage{stmaryrd}
\usepackage{caption}
\usepackage{subcaption}
\usepackage{hhline}
\usepackage{url}
\usepackage{cite}
\usepackage{hyperref}
\usepackage{enumitem}

\usepackage{epstopdf}

\usepackage{bm} %for bold greek symbols

\usepackage{ulem} % Needed for \sout

\usepackage{todonotes}

\usepackage{xcolor}
\usepackage{soul}

\graphicspath{ {./Figures/}{../} } % TODO: We cannot have a folder with figures for submission. All figures in same folder as latex files

\newcommand{\CH}[2]{\text{CH}\ensuremath{_{#1}^{#2}}}
\newcommand{\N}[2]{\text{N}\ensuremath{_{#1}^{#2}}}
\renewcommand{\H}[2]{\text{H}\ensuremath{_{#1}^{#2}}}
\renewcommand{\O}[2]{\text{O}\ensuremath{_{#1}^{#2}}}

\begin{document}

\title{3D particle simulations of positive air-methane streamers for combustion}
\author{Dennis Bouwman$^1$, Jannis Teunissen$^1$, Ute Ebert$^{1,2}$}
\address{$^1$ Centrum Wiskunde \& Informatica (CWI), Amsterdam, The Netherlands}
\address{$^2$ Department of Applied Physics, Eindhoven University of Technology, PO Box 513, 5600 MB Eindhoven, The Netherlands}

\eads{\mailto{Dennis.Bouwman@cwi.nl}}

\begin{abstract}
\noindent
Streamer discharges can be used as a primary source of reactive species for plasma-assisted combustion. In this research we investigate positive streamers in a stoichiometric air-methane mixture at $1$\,bar and $300$\,K with a three-dimensional particle-in-cell model for the electrons. We first discuss suitable electron scattering cross sections and an extension of the photoionization mechanism to air-methane mixtures. We discuss that the addition of $9.5\%$ methane leaves electron transport and reaction coefficients essentially unchanged, but it largely suppresses photoionization and shortens the photon mean free path. This leads to (1) accelerated streamer branching, (2) higher electric field enhancement at the streamer head, (3) lower internal electric fields, and (4) higher electron densities in the streamer channel. We also calculate the time-integrated energy density deposited during the evolution of positive streamers in background electric fields of 12.5 and 20~kV/cm. We find typical values of the deposited energy density in the range of $0.5-2.5$\,kJ/m$^{3}$ within the ionized interior of streamers with a length of $5$\,mm; this value is rather independent of the electric fields applied here. Finally we find that the energy deposited in the inelastic electron scattering processes mainly produces reactive nitrogen species: \N{2}{} triplet states and N, but also O and H radicals. The production of \H{2}{} and \O{2}{} singlet states also occurs albeit less pronounced. Our calculation of the primary production of reactive species can for example be used in global chemistry models.
\end{abstract}


\noindent Accepted for publication: \PSST on Apr 6, 2022
\maketitle
\ioptwocol % for two column layout

\section{Introduction}\label{sec:introduction}

Streamers are transient gas discharges consisting of growing plasma filaments with field enhancement at the tip. A review covering a wide range of investigations into streamer dynamics is presented in \cite{Nijdam2020ThePhenomena}. In industry, mainly positive streamer discharges are found in a variety of applications such as: plasma medicine \cite{Fridman2008AppliedMedicine}, industrial surface treatment \cite{Bardos2010ColdApplications} and air-pollution control \cite{Kim2004NonthermalProspects}. In particular, positive streamers in combustible mixtures are relevant for plasma-assisted combustion, as will be discussed in section \ref{sec:PAC}.\\ 

\subsection{Streamer dynamics in varying gases}
The properties of streamers are determined by the electron dynamics which in turn are governed by gas-specific photoionization, and by transport and reaction coefficients. %Thus, different gas compositions naturally yield different characteristics of the streamer. 
There are numerous investigations on streamers in different gases which illustrate this gas-dependency, for example: CO$_2$ \cite{Bagheri2020SimulationSources}, N$_2$:CH$_4$ \cite{Kohn2019}, air with artificially increased electron attachment or reduced photoionization \cite{Francisco2021, Bagheri2019}, \N{2}{}-\O{2}{} mixtures in various ratios \cite{Teunissen2016, Chen2018BranchingMixtures}, the atmospheres of Venus and Jupiter \cite{Dubrovin2010SpriteInvestigation}, and mixtures resembling the atmosphere of the primordial Earth \cite{Kohn2022StreamerEarth}.

Specific to \N{2}{}-\O{2}{} mixtures it was shown experimentally that streamers tend to branch more frequently for decreasing oxygen concentrations \cite{Nijdam2010ProbingMixtures, Briels2008PositiveLaws}. Decreasing the oxygen concentration gives different photoionization characteristics due to a longer photon mean free path, while the number of photoionization events stays essentially unchanged. In mixtures with low oxygen concentrations ($<0.2\%$) stochastic fluctuations associated with the discrete nature of photoionization then accelerate branching.

In an air-methane mixture, however, methane largely absorbs photons without ionizing, and it also quenches excited \N{2}{} that otherwise could emit photons. So the rate of ionizing photoabsorption decreases, and the photon mean free path decreases as well~\cite{Naidis2007}. As we will show, these two effects combined also enhance the stochastic fluctuations in the leading edge of the streamer. In simulations where the single-electron dynamics in the leading edge are sufficiently resolved (e.g.\ in a 3D PIC-MCC model) this will ultimately accelerate streamer branching.

\subsection{Plasma-assisted combustion}\label{sec:PAC}
In plasma-assisted combustion there is an interest in positive streamers propagating through combustible mixtures, such as air-methane \cite{Starikovskaia2006PlasmaCombustion, Starikovskiy2013, Ju2015}. In a streamer discharge one finds electrons with energies that exceed the gas temperature by orders of magnitude.
These energetic electrons produce excited nitrogen states, hydrogen- and oxygen-radicals, fuel fragments and other reactive species through collisions with neutral gas molecules. The resulting non-equilibrium distribution of reactive species is then available for plasma- and combustion-chemical processes taking place on slower time scales. These processes have been studied numerically. For instance in \cite{Naidis2007} the effect of a streamer discharge on the reduction in ignition delay was studied with an axisymmetric fluid model. In \cite{Breden2013, Breden2019} a two-dimensional cartesian fluid approach was used to investigate radical production by streamers in air-methane (without correcting the photoionization). In \cite{Takana2014} a similar model was used (2D cartesian without correcting the photoionization and also without electron attachment) to investigate the production of radicals for air-methane streamers at $10$\,bar and $600$\,K. An overview of multiscale modelling for plasma-assisted combustion is presented in \cite{Yang2017}. Finally a comparison between 0D and axisymmetric models for the simulation of air-methane streamers (without photoionization) is presented in \cite{Levko2017}. The application of low-temperature plasmas is found to have favourable effects such as: ultra-lean combustion for emission reduction \cite{Pilla2006StabilizationPlasma}, increased flame propagation speed \cite{Ombrello2010FlameO3, Ombrello2010FlameO2a1g} and flame stabilization \cite{Lacoste2013EffectFlame}. In a single-pulse discharge the generation of reactive species can, under the right conditions, lead to a reduction of ignition delay time \cite{Singleton2011Theair}.

\subsection{Content of the paper}
We simulate positive streamers in a stoichiometric air-methane mixture at $1$\,bar and $300$\,K in background electric fields of $12.5$\,kV/cm and $20$\,kV/cm in an $8$\,mm gap. Such conditions are relevant for the initial stages of plasma-assisted ignition. We will analyze the simulations from two different viewpoints. First we will analyze how the addition of methane affects the fundamental properties of the discharge, such as streamer branching, electric field enhancement and electron densities and energies. Second, we analyze the streamer discharge within the context of plasma-assisted combustion. We will study the deposited energy density and the G-values, i.e.\ the efficiency with which primary reactive species are produced. The production of these primary species can represent an initial condition of plasma-chemical and ignition-chemical calculations.\\

In section \ref{sec:method} we describe the particle-in-cell model, cross section sets, simulation conditions and we correct for the suppressing influence of \CH{4}{} on the photoionization mechanism. In section \ref{sec:morphology} we compare the dynamic properties of a positive streamer in air and air-methane. Finally, in section \ref{sec:kinetics} we address the plasma-chemical activation of a stoichiometric air-methane mixture by a positive streamer.

\section{Simulation Method}\label{sec:method}
We simulate a positive streamer discharge using a 3D Particle-In-Cell model with Monte-Carlo Collisions (PIC-MCC). Our implementation is based on the model described in \cite{Teunissen2016}. Here we present a short summary and describe the photoionization model used for discharges in air-methane mixtures.

\subsection{Description of PIC-MCC model}
Within a PIC-MCC model electrons are represented by super-particles. One super-particle represents a variable number of physical electrons, represented by the weight $w$ as is further elaborated in section \ref{sec:particle_weight}. The motion of electrons is then governed by an acceleration due to the electric field combined with isotropic scattering processes due to collisions with the gas molecules. Ions are included as a density and are assumed to be immobile on the considered short streamer time scale. The neutral gas particles are taken as a homogeneous background density with which the electrons can stochastically collide. Electron collisions with ions, excited/dissociated molecules and other electrons can be omitted since the ionization degree is low, around $10^{-4}$ to $10^{-5}$. \\

An advantage of a PIC-MCC model over the conventional fluid models is that the electron energy distribution function (EEDF) is approximated explicitly and without the need for assumptions such as the local field approximation. For plasma-chemical streamer applications a good approximation of electron energies is important as this quantity determines the production rate of reactive species. Another advantage of using PIC-MCC models is that it is better equipped to deal with single-electron fluctuations. In air, it is known that electron density fluctuations, in particular due to stochastic photoionization, accelerate streamer branching \cite{Bagheri2019, Luque2011ElectronAir, Wang2022AAir}. In the case of air-methane mixtures there is less photoionization and the photon mean free path is shorter (see section \ref{sec:photoionization}). As a result single-electron fluctuations occur in the active zone, i.e. the region where the electric field is above breakdown. In order to properly resolve the influence of these single-electron fluctuations we perform simulations with the PIC-MCC model. \\

The main drawback of PIC-MCC models is a high computational cost associated with the use of a large number of particles, especially in three dimensions. For example, typical computing times for our simulations are on the order of days (performed on one node of Cartesius, the Dutch national supercomputer), whereas a two-dimensional fluid simulation performed on an `average' desktop typically only requires several minutes of computation. More details about such a comparison are found in the appendix of \cite{Wang2022AAir}.

\subsection{Particle weight}\label{sec:particle_weight}

In a streamer discharge the number of free electrons increases rapidly to the point where it becomes computationally infeasible to simulate every electron individually. To overcome this limitation the weight (\mbox{$w\geq 1$}) of a computer particle is dynamically updated, thereby allowing one particle to represent one or many physical electrons. This technique, called \textit{adaptive particle management}, ensures that the computational complexity of the simulation remains tractable at the cost of an artificial noise on the electron density.  Details on the implementation and performance of this algorithm can be found in \cite{Teunissen2014ControllingTrees}. The central idea is that particles are merged and/or split between time steps, thereby changing their weights in order to bring them close to a desired weight $w_d$:
\begin{equation}
    w_d = \frac{n_e \Delta x^3}{N_{\text{ppc}}},
\end{equation}
where $n_e$ is the local electron density, $\Delta x^3$ is the volume of the cell containing the particle and $N_{\text{ppc}}$ is the desired number of particles per cell, which we have chosen at $100$. 

Now we focus on the influence of artificial noise introduced by super-particles before estimating that for our parameters this effect is small. To that end we note that electron density fluctuations accelerate branching. This was shown, for example, in \cite{Luque2011ElectronAir} using a stochastic fluid model. A similar conclusion was drawn in \cite{Bagheri2019} and later in \cite{Wang2022AAir} by combining a conventional fluid model with a stochastic version of the Zhelezniak photoionization model. Therefore with super-particles (with \mbox{$w>1$}) it is inevitable that electron density fluctuations are more prevalent compared to using only single particles (\mbox{$w=1$}). As a result streamer branching would occur more often, especially when compared with fluid models which neglect physical density fluctuations. However, by choosing a small cell volume and a high $N_{\text{ppc}}$ one can ensure that the artificial noise introduced by super-particles does not dominate the fluctuations introduced by physical mechanisms in the leading edge (such as photoionization). For example, in our simulations the smallest cell volume, which is used in the high-field region at the streamer head, equals \mbox{$(4.0\,\mu$}m$)^3$. Thus electron avalanches on the finest grid with densities below \mbox{$1.6\cdot 10^{18}$}\,m$^{-3}$ are simulated using particles with unit weight. In section \ref{sec:branching} we show that this is sufficiently accurate in order to resolve the single-electron fluctuations in the leading edge.

\subsection{Cross sections for electron collisions} \label{sec:cross sections}
A set of cross sections for the dominant scattering processes is required to describe the electron kinetics. Many of the available sets have adjusted individual cross sections in order to ensure that electron transport and reaction coefficients are correctly reproduced in numerical swarm experiments. A downside of this procedure is that adjusting cross sections can lead to incorrect reaction rates. Thus the swarm-fitting procedure is limited by non-uniqueness \cite{Petrovic2007}, since different modifications to inelastic cross sections can result in the same swarm parameters but with different reaction rates. In earlier work, we defined an unfitted cross section set and addressed the issue of non-uniqueness for the case of \CH{4}{} \cite{Bouwman2021NeutralSet}.\\

Given our focus on accurately predicting the produced reactive species we only use unfitted cross section sets. For \N{2}{} we adopt the cross section set of Kawaguchi~\textit{et al.}~\cite{Kawaguchi2021Electron/Ar}, but neglect the inter-rotational (de-)excitation which is only relevant at low reduced electric fields. For \O{2}{} we adopt the cross sections recommended by Itikawa~\textit{et al.}~\cite{Itikawa2009}. Electron attachment by three-body collisions with \O{2}{} are taken from \cite{Lawton1978ExcitationElectrons}. For \CH{4}{} we adopt the cross section set proposed by Bouwman~\textit{et al.}~\cite{Bouwman2021NeutralSet} which are based on the recommendations by Song~\textit{et al.}~\cite{Song2017} combined with cross sections for the neutral dissociation processes. Finally, all scattering processes are assumed to be isotropic.\\

\begin{figure}
    \centering
    \includegraphics[width=0.5\textwidth]{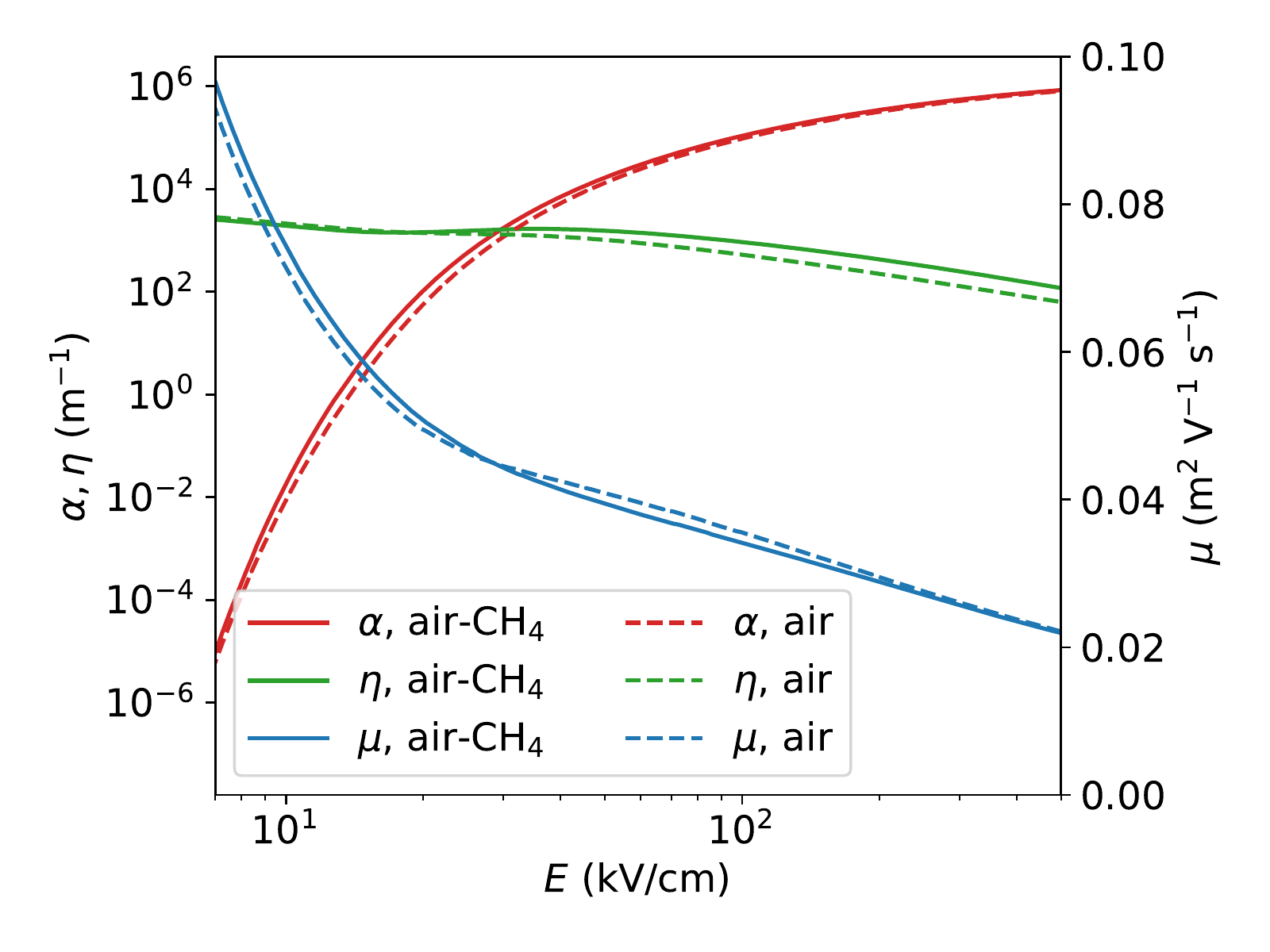}
    \caption{The transport and reaction coefficients for air and stoichiometric air-methane at $1$\,bar and \mbox{$300$\,K}. The axis for the mobility $\mu$ is linear, whereas the axis for ionization and attachment coefficients $\alpha$ and $\eta$, respectively, is logarithmic. We conclude that a gas fraction of $9.5\%$ \CH{4}{} introduces only minor changes to these coefficients.}
    \label{fig:swarm_parameters}
\end{figure}

For a stoichiometric air-methane mixture ($71.5\%$ \N{2}{}, $19\%$ \O{2}{} and $9.5\%$ \CH{4}{}) at $300$\,K and $1$\,bar we compared the ionization and attachment coefficients $\alpha,\,\eta$ and the electron mobility $\mu$ with air, in figure \ref{fig:swarm_parameters}. These coefficients are calculated using BOLSIG+ (desktop version of 2019) using the temporal growth setting and assuming isotropic scattering \cite{Hagelaar2005, BOLSIG+}. This relatively small fraction of \CH{4}{} introduces only minor quantitative differences. Furthermore, we remind the reader that the coefficients depicted here are only used for the purpose of illustration. Within a PIC-MCC model the electron kinetics is directly determined by the cross sections. 

\subsection{Photoionization in air-methane mixtures}\label{sec:photoionization}
Naidis \cite{Naidis2007} approximated the influence of \CH{4}{} on photoionization in air-methane by including an extra absorption factor in the classical Zhelezniak photoionization model \cite{Zhelezniak1982PhotoionizationDischarge}. In that work, however, no corrections were made for the changing effective quenching pressure due to the addition of methane. An alternative to the Zhelezniak photoionization model (in air) is discussed in \cite{Marskar20203DPhotoionization, Stephens2018PracticalTransport}. Here we follow the reasoning of \cite{Naidis2007} as we  extend the Monte-Carlo Zhelezniak photoionization model, such as presented in \cite{Teunissen2016,Chanrion2008AAir}, to air-methane mixtures. We formulate necessary corrections to include photon-loss due to non-ionizing photoabsorption and, notably, quenching. \\

\subsubsection{Quenching: }
In dry air, it is well-established that radiative transitions of excited states of \N{2}{} emit photons in the energy range that is associated with the photoionization of \O{2}{}, namely $12.1-12.65$\,eV. On the contrary, the excited states of \CH{4}{} are all dissociative \cite{Song2017} and do not emit photons with energies that can ionize \O{2}{}. Thus, photons in the relevant energy range are produced in air-methane mixtures only by the excited states of \N{2}{}. However, \CH{4}{} does suppress the total number of photons by quenching the excited states of \N{2}{}. The effective quenching pressure $p_q^{\text{eff}}$ can be written as:
\begin{equation}
    \frac{1}{p_q^{\text{eff}}} = \frac{\chi^{\O{2}{}}}{p_q^{\O{2}{}}} + \frac{\chi^{\N{2}{}}}{p_q^{\N{2}{}}} + \frac{\chi^{\CH{4}{}}}{p_q^{\CH{4}{}}},
\end{equation}
where $\chi$ denotes the respective gas number fraction. The quenching pressures are reported in \cite{Albugues1974} to be: ${p_q^{\O{2}{}}=3.8}$\,torr, ${p_q^{\N{2}{}}=91}$\,torr and ${p_q^{\CH{4}{}}=1.8}$\,torr. According to the Zhelezniak model, the average number of photons $\bar\eta$ produced per impact ionization of an air molecule (i.e.\ \N{2}{} or \O{2}{}) is given by:
\begin{equation}
    \bar\eta = \frac{p_q^{\text{eff}}}{p_q^{\text{eff}} + p}\ \xi,
\end{equation}
with the pressure $p$ and the factor $\xi$ relating the radiative de-excitation rate of \N{2}{} to the ionization rate. The dependence of $\xi$ on the local electric field is only partially tabulated in \cite{Zhelezniak1982PhotoionizationDischarge}, so we have for simplicity taken the constant value \mbox{$\xi=0.05$}. This value is within the ranges considered in \cite{Bagheri2019, Li2021ComparingValidation}, where it was shown that that deviations by a factor two have little influence on most streamer properties, although lower values can increase the probability of streamer branching.

\begin{figure}
    \centering
    \includegraphics[width=0.5\textwidth]{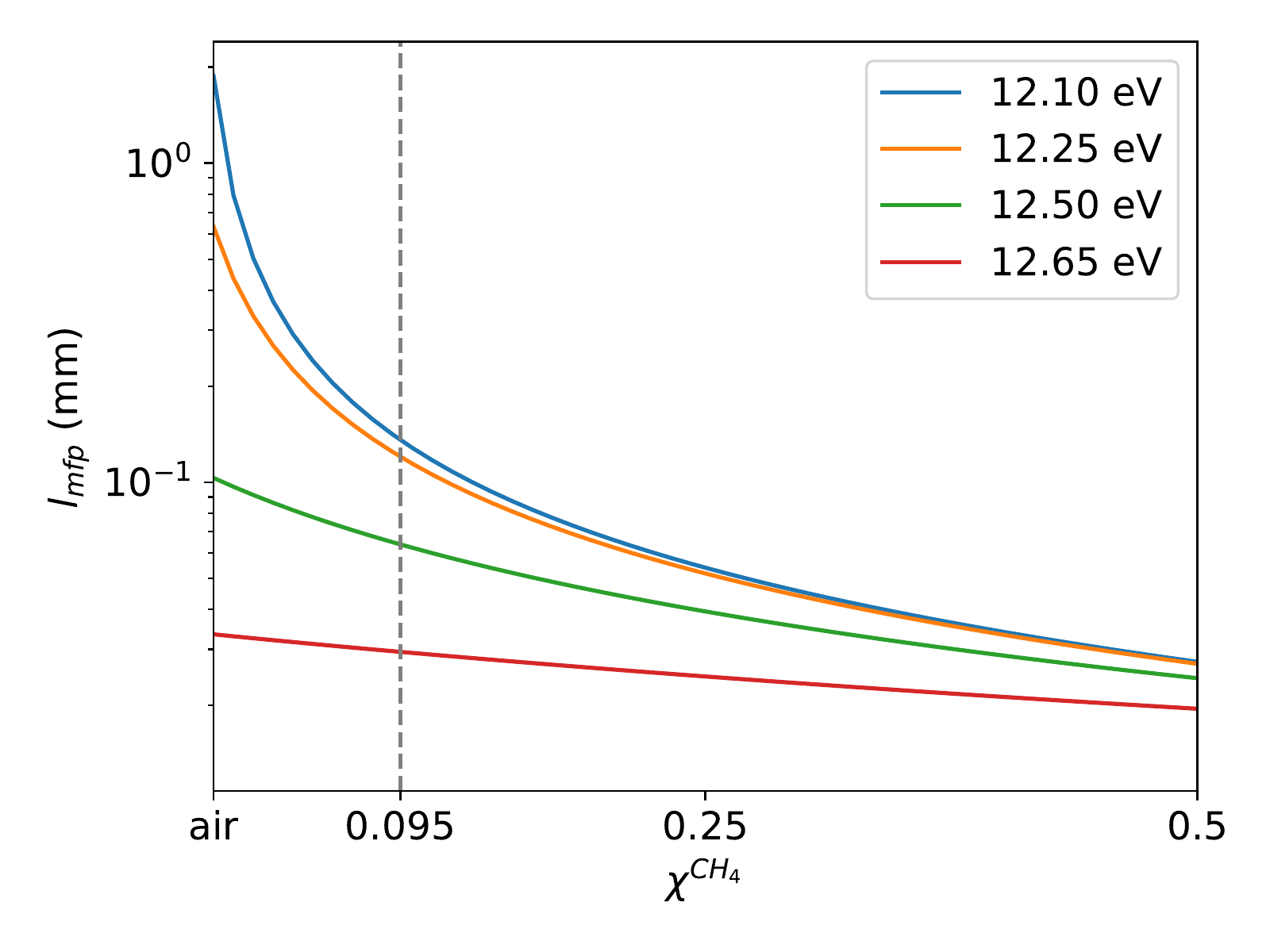}
    \caption{The photon mean free path as a function of the CH$_4$ fraction $\chi^{\rm CH_4}$ in air-methane mixtures for various photon energies. Stoichiometric conditions are denoted by the dashed grey line. Adding methane shortens the photon mean free path.}
    \label{fig:photon_meanfreepath}
\end{figure}

\subsubsection{Photoabsorption: }
The absorption of photons is determined by the photoabsorption cross sections $\sigma$. Kameta~\textit{et al.}~\cite{Kameta2002} have determined the photoabsorption cross sections for \CH{4}{}. In the energy range relevant for the ionization of oxygen, \mbox{$12.1-12.65$}\,eV, they found that the cross section for photoionization of \CH{4}{} is negligble compared to non-ionizing photoabsorption of \CH{4}{} (and also to the photoionization of \O{2}{}).  Furthermore, in this energy region the value of $\sigma^{\CH{4}{}}$ is roughly constant at $3.0\cdot 10^{-17}$\,cm$^2$. On top of that, we only have to consider the cross section for ionizing photoabsorption of \O{2}{} in this energy range.  $\sigma^{\O{2}{}}$ is typically described as a power function of photon energy $\gamma$, whereas the cross sections for \N{2}{} can be neglected \cite{Pancheshnyi2014Photoionization2}. This allows us to write the mean free path of a photon with an energy $\gamma$ in interval \mbox{$12.1-12.65$}\,eV as:      
\begin{equation}
    l_{\text{mfp}}(\gamma)=\left(\sigma^{\CH{4}{}}n^{\CH{4}{}}+\sigma^{\O{2}{}}(\gamma)n^{\O{2}{}}\right)^{-1},
\end{equation}
with $n$ representing the number density of a gas component (the number density $N$ of an ideal gas is \mbox{$2.4\cdot 10^{25}$}\,m$^{-3}$ at $1$\,bar and $300$\,K). The mean free path plays an important role in the dynamics of electron density fluctuations which are a result of stochastic photoionization. For that reason we have illustrated the dependency of $l_{\rm mfp}$ on the gas fraction of \CH{4}{} in figure \ref{fig:photon_meanfreepath}. The gas composition was determined by keeping the ratio between \N{2}{} and \O{2}{} fixed at \mbox{$79:21$} while varying the fraction of \CH{4}{}. \\

Finally, we formulate the probability of a photon to ionize \O{2}{} (as opposed to being lost due to absorption by \CH{4}{}):
\begin{equation}
    P(\gamma) = \frac{\sigma^{\O{2}{}}(\gamma)n^{\O{2}{}}}{\sigma^{\CH{4}{}}n^{\CH{4}{}}+\sigma^{\O{2}{}}(\gamma)n^{\O{2}{}}}.
\end{equation}

\subsubsection{Implementation: } 
The photoionization procedure in air-methane mixtures is implemented as follows: if a super-particle with weight $w$ ionizes an \N{2}{} or \O{2}{} molecule (note that our choice for the parameter $\xi$ in the Zhelezniak model corresponds to air), then that produces a random number of photons which are sampled from a Poisson distribution with mean $\bar\eta w$. Each of these photons are produced individually, i.e.\ the use of super-photons is excluded, and are assigned a random energy $\gamma$. Since the energy interval for ionization of \O{2}{} is assumed to be uniformly populated, we take $\gamma$ as a uniform random variable in the interval \mbox{$12.1-12.65$}\,eV.  Then a Bernoulli trial with probability $P(\gamma)$ determines whether the photon is lost due to absorption by \CH{4}{}. If not, the photon is emitted isotropically with a travel distance drawn from a Poisson distribution with mean $l_{\text{mfp}}(\gamma)$ upon which it ionizes an \O{2}{} molecule.\\

\subsubsection{Interpretation: }
\begin{figure}
    \centering
    \includegraphics[width=0.5\textwidth]{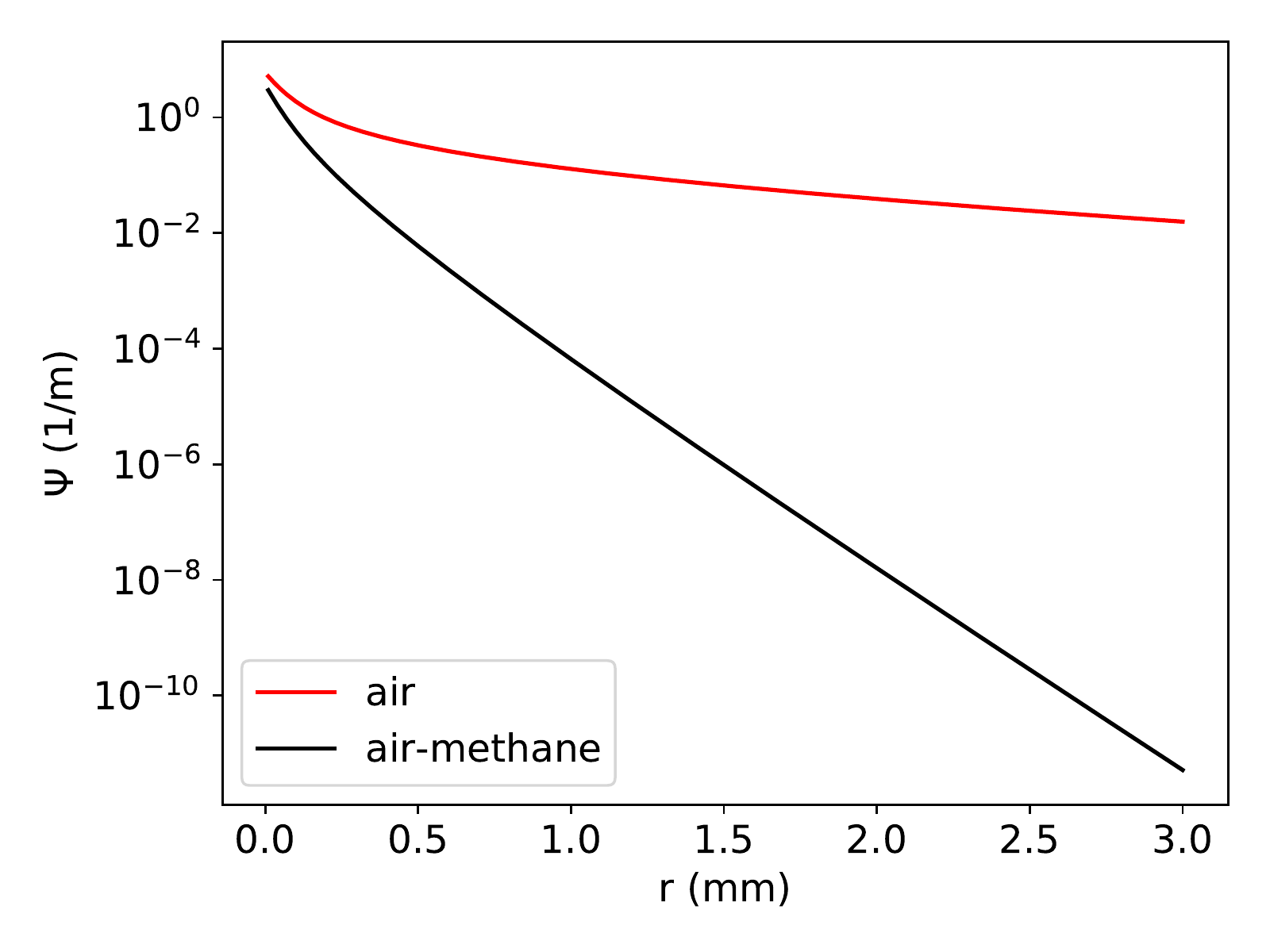}
    \caption{The absorption function $\Psi$ for air and an air-methane mixture containing $9.5\%$ methane. This function illustrates the damping influence of methane on the photoionization mechanism.}
    \label{fig:absorption_function.pdf}
\end{figure}
Now we illustrate the influence of \CH{4}{} on photoionization. We do this by computing the absorption function $\Psi$, which can be expressed by the absorption function in air, $\Psi_{\rm air}$ as:
\begin{align}\label{eq:psi}
    \Psi &=e^{-r\mu^{\CH{4}{}}}\Psi_{\rm air},\\
    &=e^{-r\mu^{\CH{4}{}}}\cdot\frac{e^{-r\mu_{\rm min}}-e^{-r\mu_{\rm max}}}{r \ln(\mu_{\rm max}/\mu_{\rm min})}, 
\end{align}
with $\mu^{\CH{4}{}}$ the (constant) absorption coefficient due to methane, and $\mu_{\rm min}$ and $\mu_{\rm max}$ the absorption coefficients of dry air at $300$\,K and $1$\,bar according to \cite{Zhelezniak1982PhotoionizationDischarge, Pancheshnyi2014Photoionization2}.
This function has been calculated for air and an $9.5\%$ air-methane mixture (with \mbox{$\N{2}{}:\O{2}{}$} as \mbox{$79:21$}) and shown in figure \ref{fig:absorption_function.pdf}. Note that for the purpose of this illustration we have assumed that $\xi$ is constant for both mixtures. Clearly, the addition of methane leads to a strong decay of the absorption function on the millimeter length scale. This length scale is relevant for the leading edge dynamics of the streamers considered in this research. Note that $\Psi$ is not directly used in the PIC-MCC model, since photoionization events follow from sampling of relevant distributions as described in the previous section. Here, $\Psi$ is only used for the purpose of illustration.

\subsection{Computational domain and initial conditions}\label{sec:parameters}
The simulated domain consists of a cube with a length of $10$\,mm for each Cartesian coordinate. The domain is filled with air or an air-methane mixture, consisting of $71.5\%$ \N{2}{}, $19\%$ \O{2}{} and $9.5\%$ \CH{4}{}, at $300$\,K and $1$\,bar. Such number fractions correspond to stoichiometric burning conditions of methane (i.e.\ \mbox{$\CH{4}{}:\O{2}{}$} as \mbox{$1:2$} and \mbox{$\N{2}{}:\O{2}{}$} as \mbox{$79:21$}).\\

We are considering a plate-to-plate geometry with a grounded plate at the bottom of the domain and a high-voltage plate at the top. Furthermore the high-voltage electrode contains an axisymmetric protrusion with length of $1.8$\,mm and a radius of $200$\,$\mu$m. The tip of this needle-electrode is a hemisphere with the same radius, giving the electrode a total length of $2$\,mm. To solve for the electrostatic potential, we use the multigrid solver described in \cite{Teunissen2018Afivo:Methods} which was recently generalized to include irregular boundaries. On the top electrode (including protrusion) a constant voltage $\phi$ of $12.5$\,kV or $20$\,kV is applied. Due to field enhancement near the needle the electric field is locally above breakdown. Far away from the needle the field relaxes to $12.5$\,kV/cm and $20$\,kV/cm respectively, which we will refer to as the background field $E_0$. The electric field is calculated by applying Dirichlet boundary conditions for the electric potential on the electrodes. The boundary conditions of the electric potential on the sides of the domain are given by homogeneous Neumann conditions, i.e.\ the field is parallel to the boundary. Moreover, super-particles are removed from the simulation if they are transported into the needle-electrode or out of the domain. Furthermore, the numerical grid is provided by the Afivo-framework \cite{Teunissen2018Afivo:Methods} which utilizes adaptive mesh-refinement (AMR) with a minimum cell size of $4.0$\,$\mu$m. This cell size is sufficiently small in order to resolve the dynamics in the thin space-charge layer. Moreover, in \cite{Wang2022AAir} it was found that the particle model is less sensitive to the cell size than a fluid model, at least when comparing streamer velocities.\\

\begin{figure}
    \centering
    \includegraphics[width=0.49\textwidth]{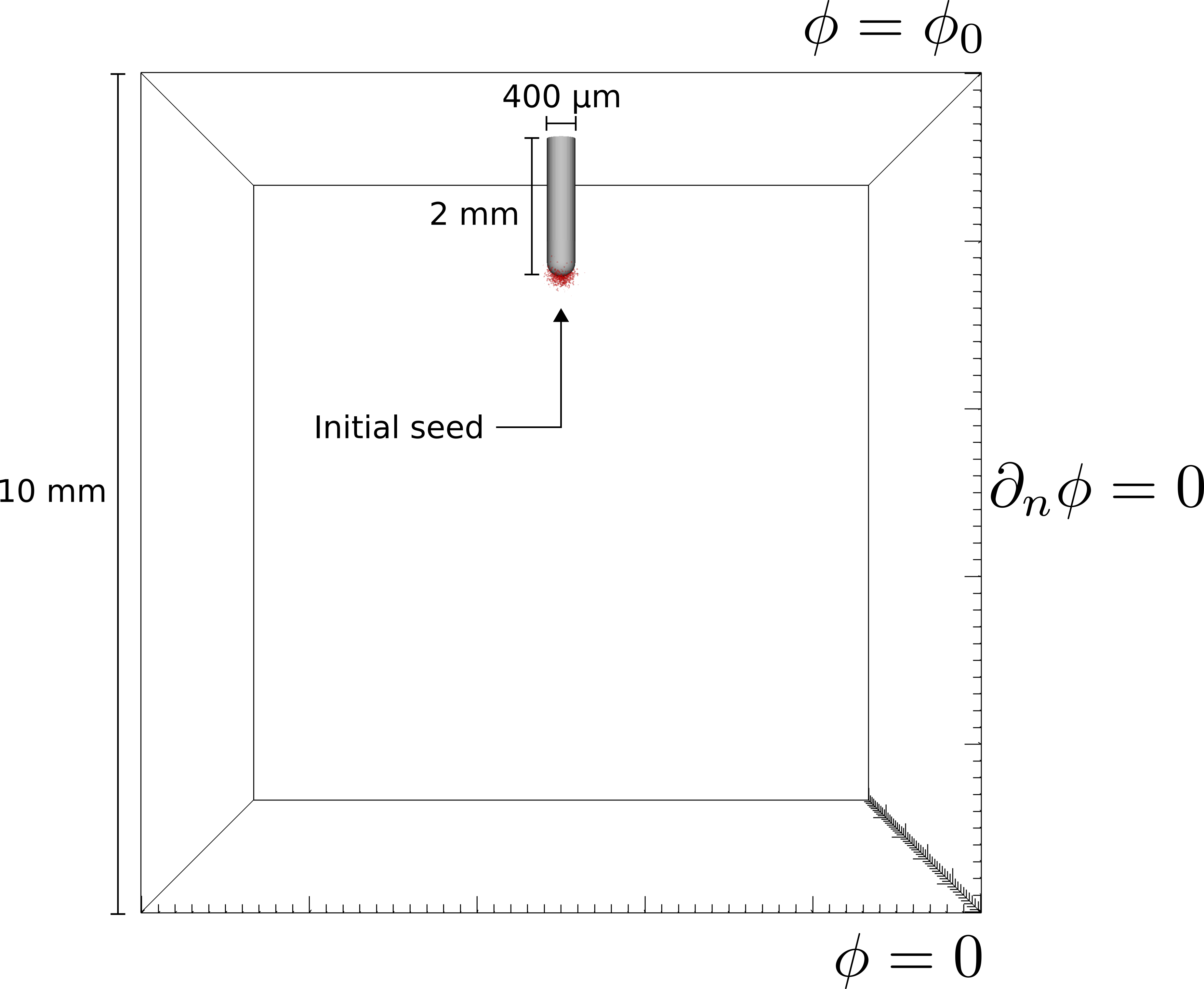}
    \caption{The computational domain consists of a Cartesian cube with a length of $10$\,mm in each coordinate. The initial seed electrons are shown in red.}
    \label{fig:simulation_domain}
\end{figure}

For all simulations in this work we use the same initial conditions consisting of a neutral seed around the electron tip. The seed consists of $1000$ electron-ion pairs at coordinates that are drawn from a Gaussian distribution centered at the tip of the electrode with a variance of $125\,\mu$m$^2$. Coordinates coinciding with the interior of the needle-electrode are rejected. The electrons are represented by particles with unit weight whereas the ions are represented as a density. Such a neutral seed is convenient when comparing discharges under different conditions as it leads to fast inception which is not highly stochastic (studying inception is not the focus of this work). The electron density of the initial seed is illustrated in figure \ref{fig:simulation_domain} in red. \\

\section{Comparison of air and air-methane streamers} \label{sec:morphology}
Here we study how the addition of methane changes streamer properties. We focus on branching, electric field enhancement and electron energies.
\subsection{Streamer branching}\label{sec:branching}

\begin{figure*}
\rotatebox{90}{\vphantom{A}} 
\begin{minipage}[t]{0.15\textwidth}
\center{$L_z=0.1$\,mm\par\medskip}
\end{minipage}
\begin{minipage}[t]{0.15\textwidth}
\center{$1.1$\,mm\par\medskip}
\end{minipage}
\begin{minipage}[t]{0.15\textwidth}
\center{$2.1$\,mm\par\medskip}
\end{minipage}
\begin{minipage}[t]{0.15\textwidth}
\center{$3.1$\,mm\par\medskip}
\end{minipage}
\begin{minipage}[t]{0.15\textwidth}
\center{$4.1$\,mm\par\medskip}
\end{minipage}
\begin{minipage}[t]{0.15\textwidth}
\center{$5.1$\,mm\par\medskip}
\end{minipage}

\rule[1ex]{\textwidth}{0.5pt}

\rotatebox{90}{\qquad \qquad \qquad Air}
\begin{minipage}[t]{0.15\textwidth}
    \includegraphics[trim=1200 1050 1200 562.5, clip, width=\textwidth]{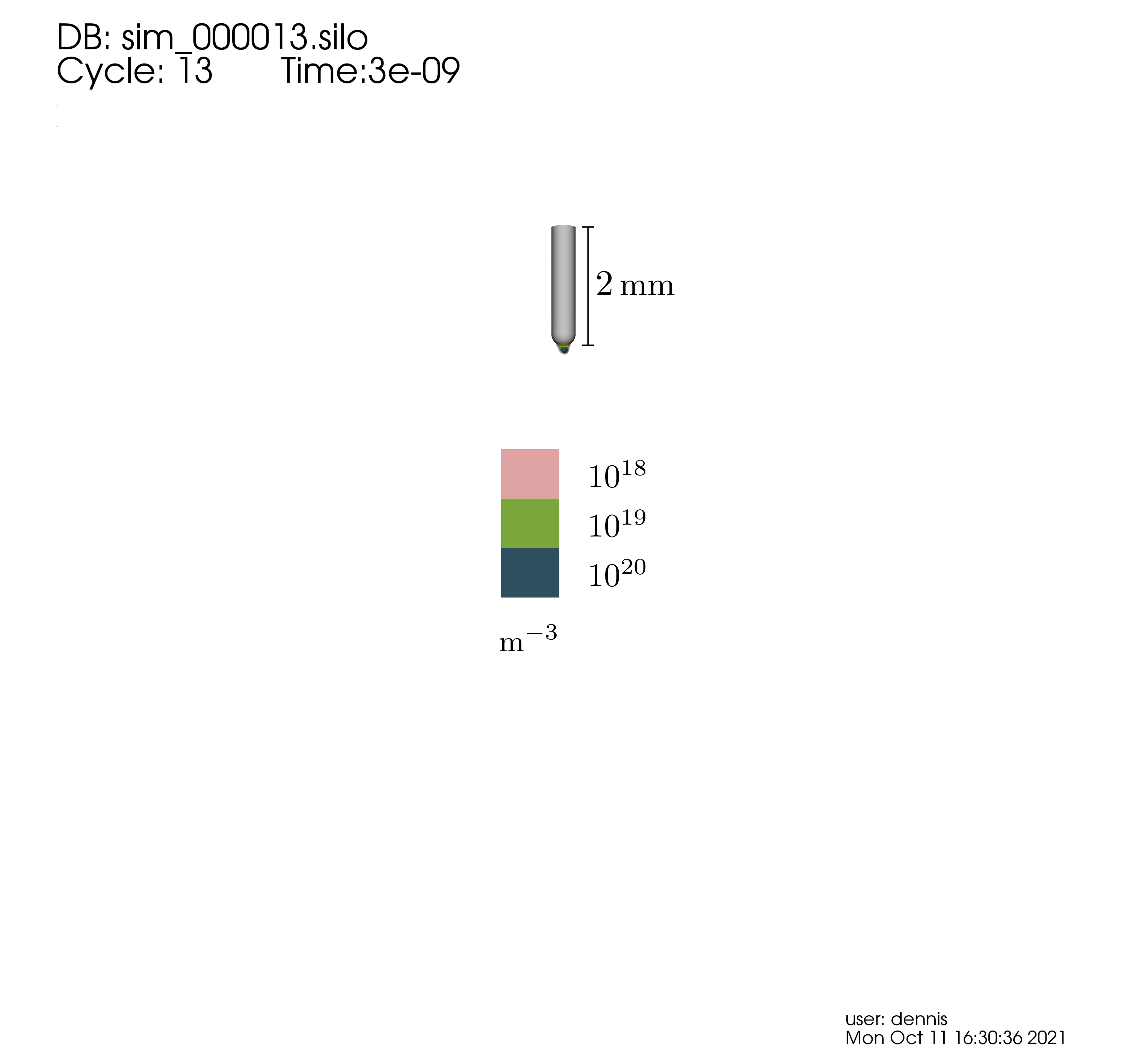}
    % This (inkscape-annotated) image has different dpi and thus needs to be trimmed with a ratio 3:4 to align properly
    \subcaption{$3.00$\,ns}
\end{minipage}    
\begin{minipage}[t]{0.15\textwidth}
\centering
    \includegraphics[trim=1600 1400 1600 750, clip, width=\textwidth]{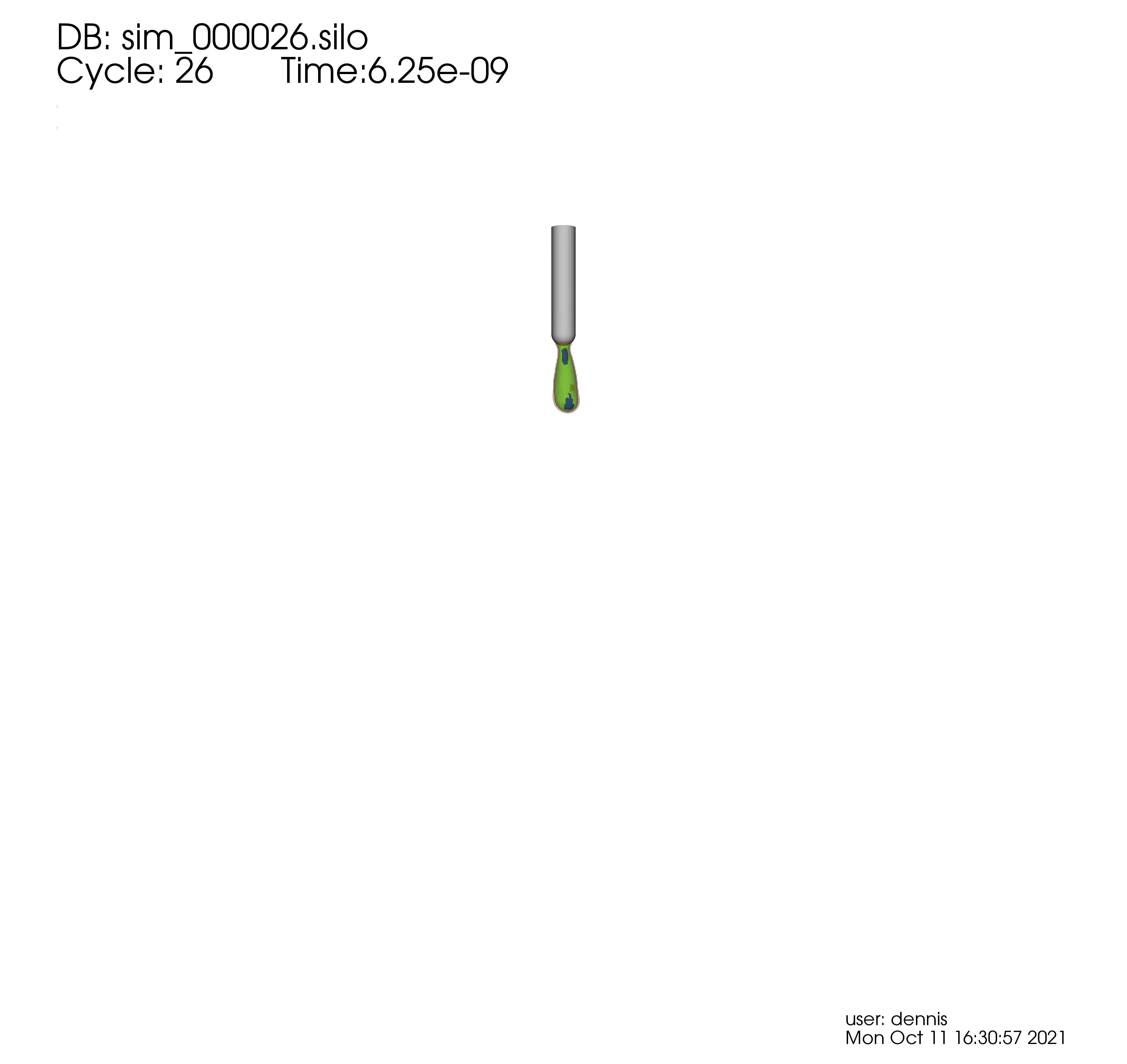}
    \subcaption{$6.25$\,ns}
\end{minipage}
    \begin{minipage}[t]{0.15\textwidth}    
    \centering
    \includegraphics[trim=1600 1400 1600 750, clip, width=\textwidth]{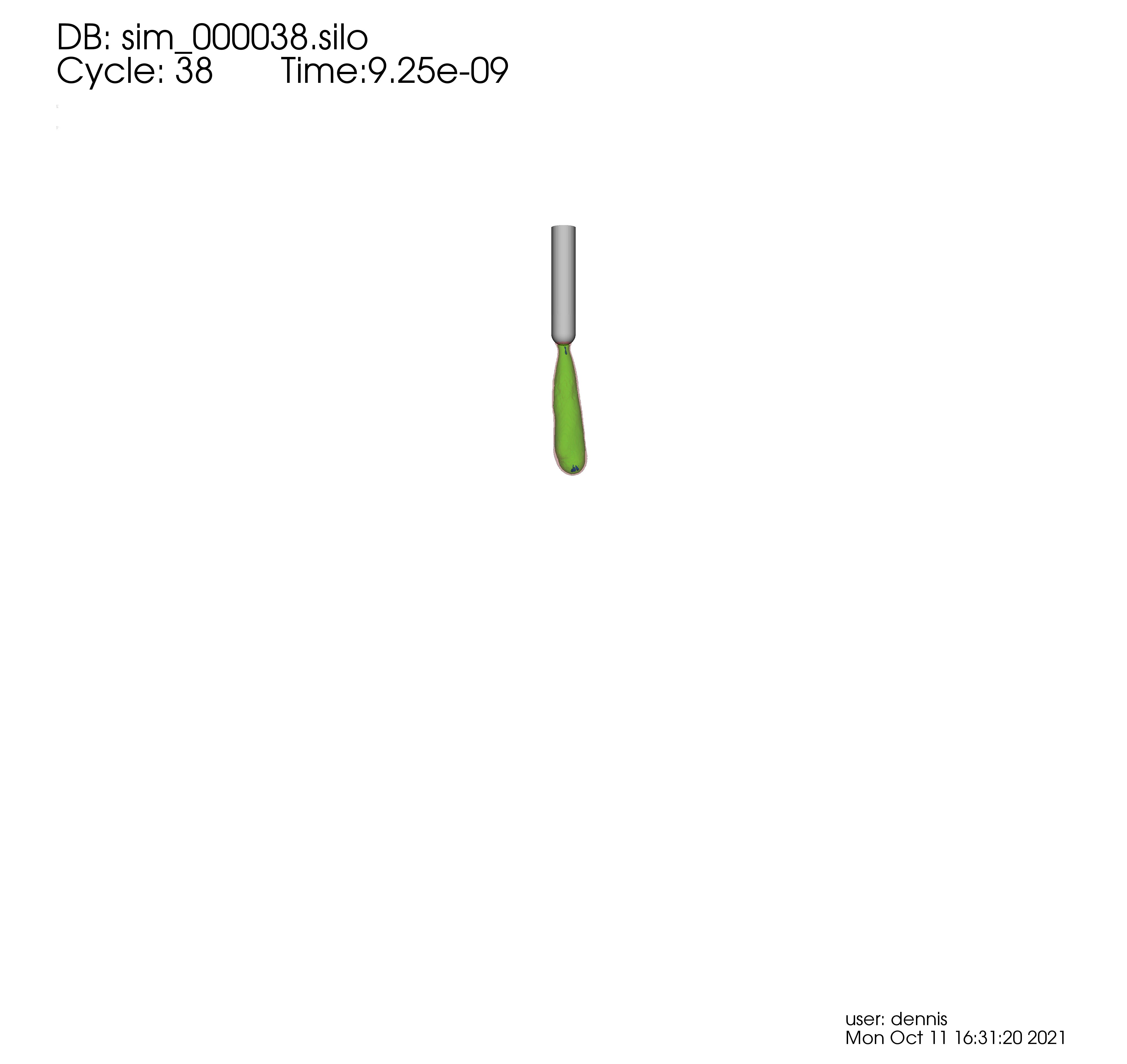}
    \subcaption{$9.25$\,ns}
\end{minipage}
\begin{minipage}[t]{0.15\textwidth}    
    \centering
    \includegraphics[trim=1600 1400 1600 750, clip, width=\textwidth]{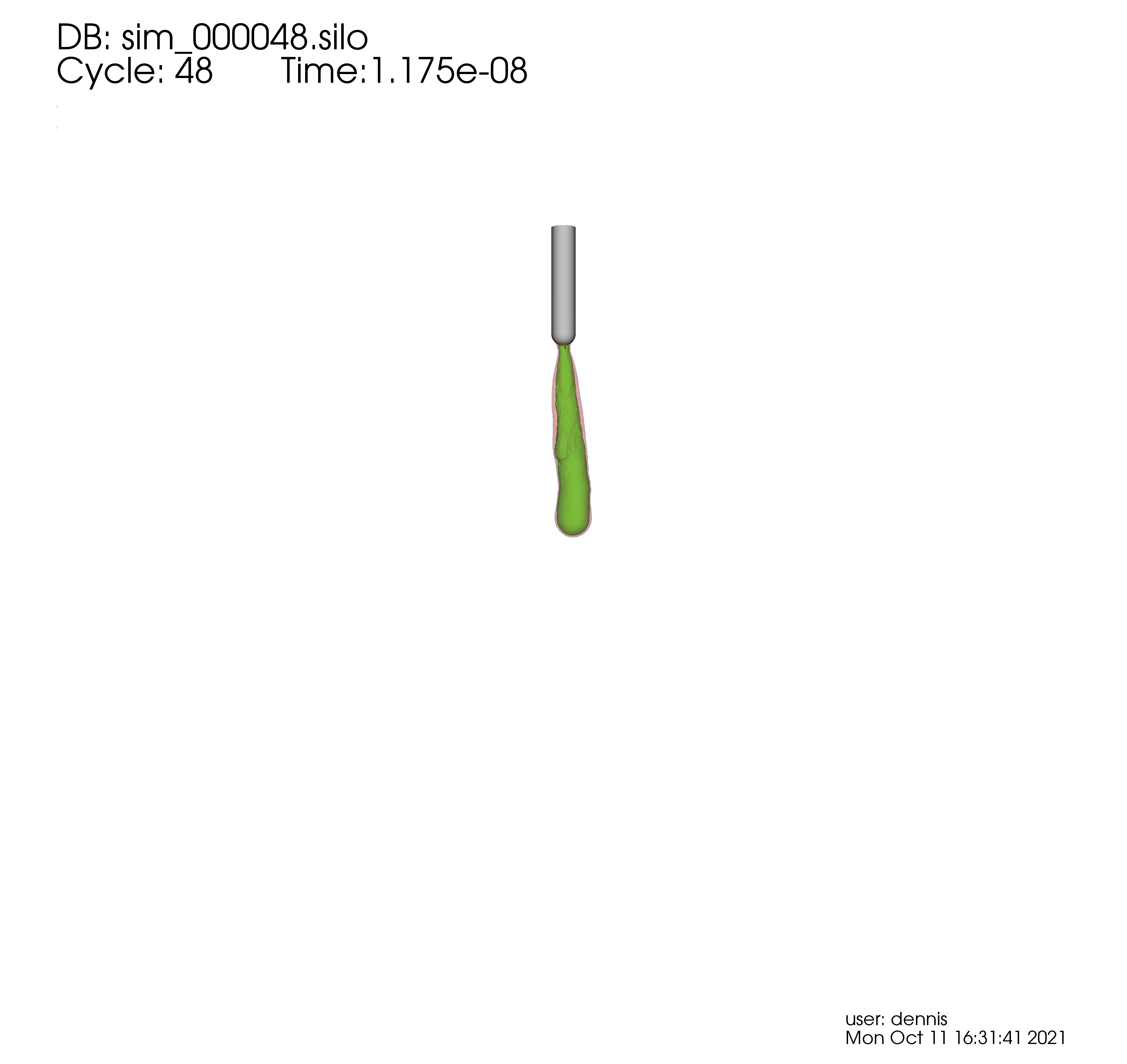}
    \subcaption{$11.75$\,ns}
\end{minipage}
\begin{minipage}[t]{0.15\textwidth}    
    \centering
    \includegraphics[trim=1600 1400 1600 750, clip, width=\textwidth]{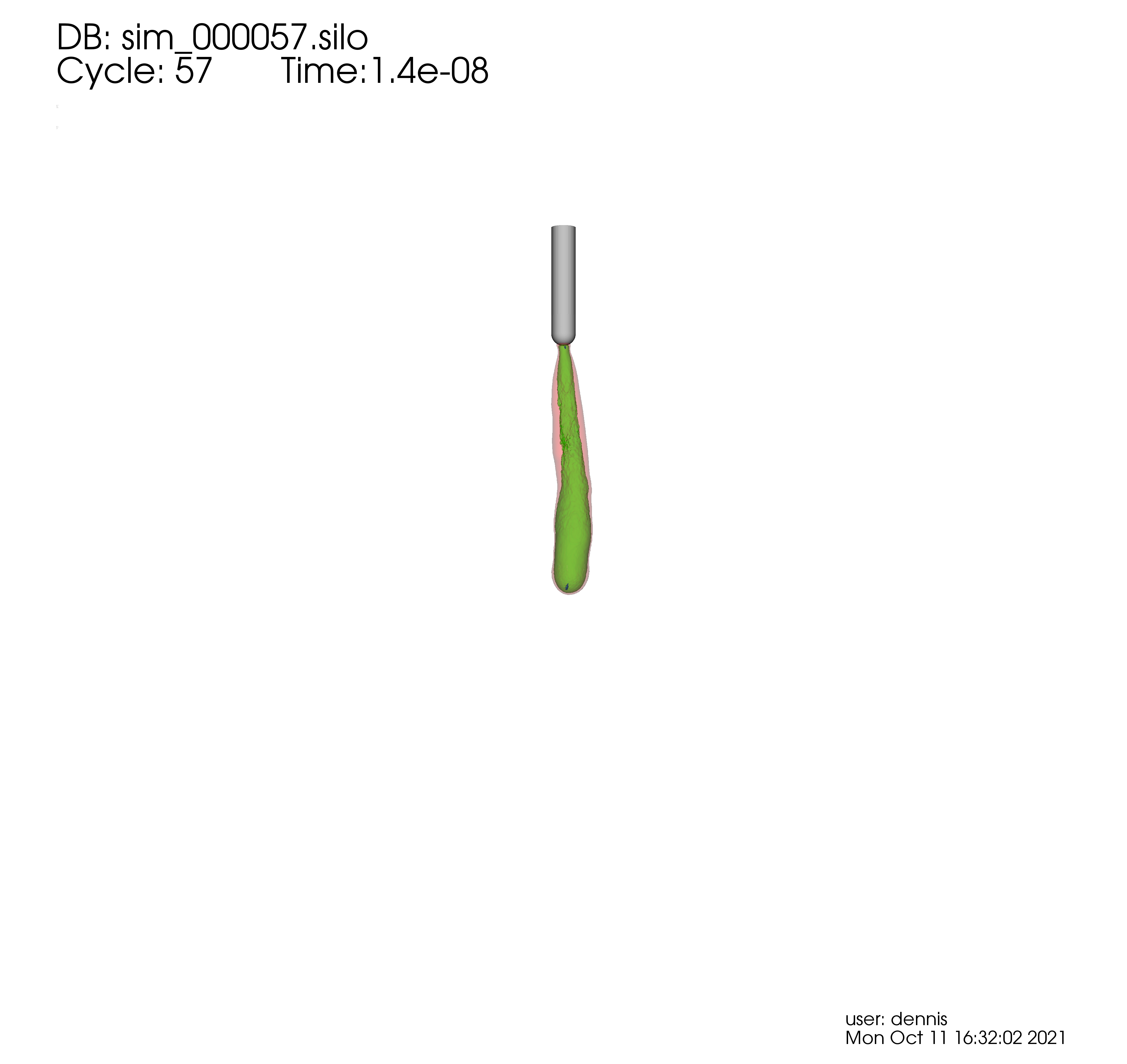}
    \subcaption{$14.00$\,ns}
    \label{subfig:air_5}
\end{minipage}
\begin{minipage}[t]{0.15\textwidth}    
    \centering
    \includegraphics[trim=1600 1400 1600 750, clip, width=\textwidth]{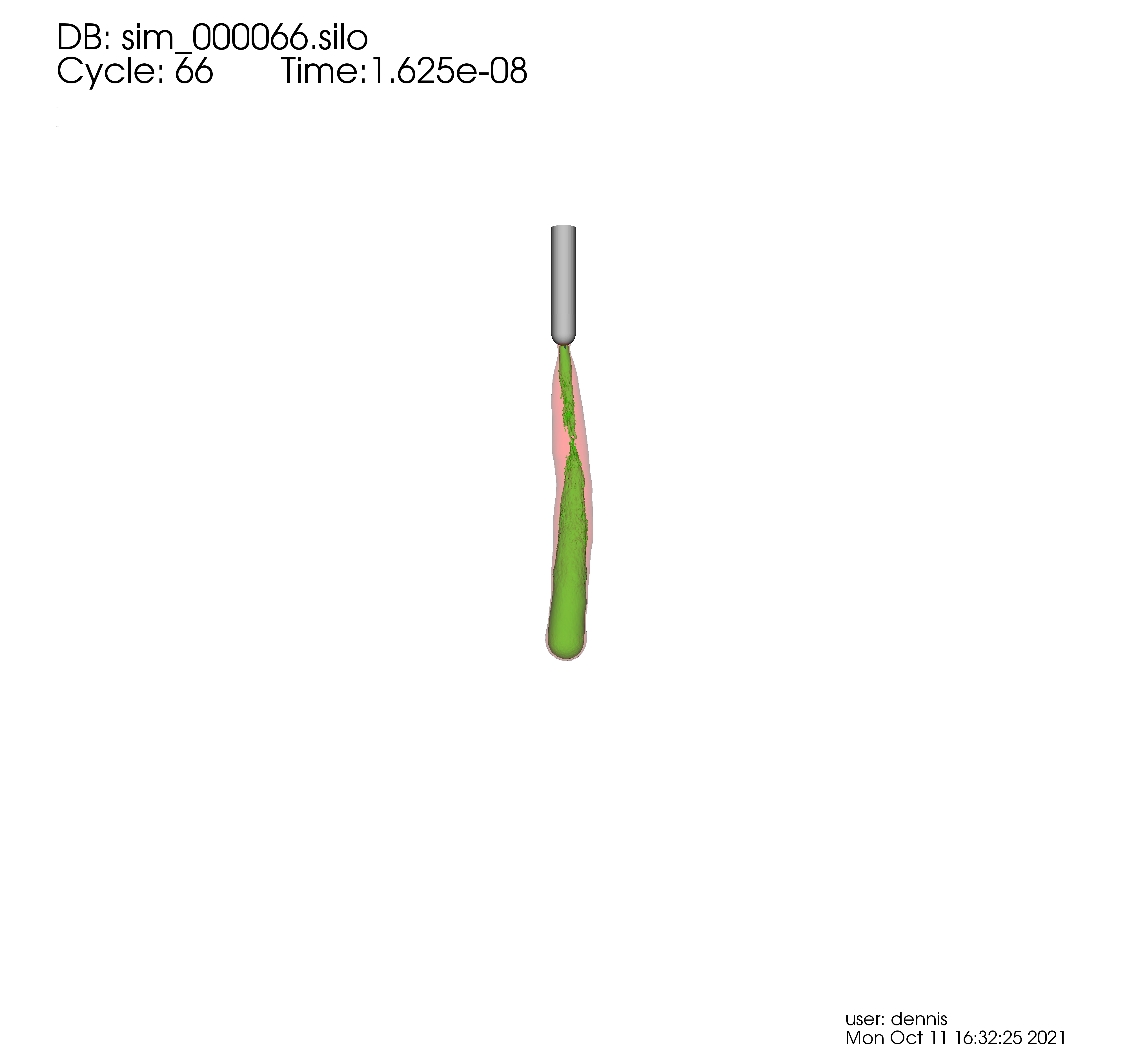}
    \subcaption{$16.25$\,ns}
    \label{subfig:air_6}
\end{minipage}

\bigskip

\rotatebox{90}{\qquad \qquad Air-methane}
\begin{minipage}[t]{0.15\textwidth}
    % {\includegraphics[trim=1200 1050 1200 562.5, clip, width=\textwidth]{v2_air_methane/g4275.png}}
    % % This (inkscape-annotated) image has different dpi and thus needs to be trimmed with a ratio 3:4 to align properly
    {\includegraphics[trim=1600 1400 1600 750, clip, width=\textwidth]{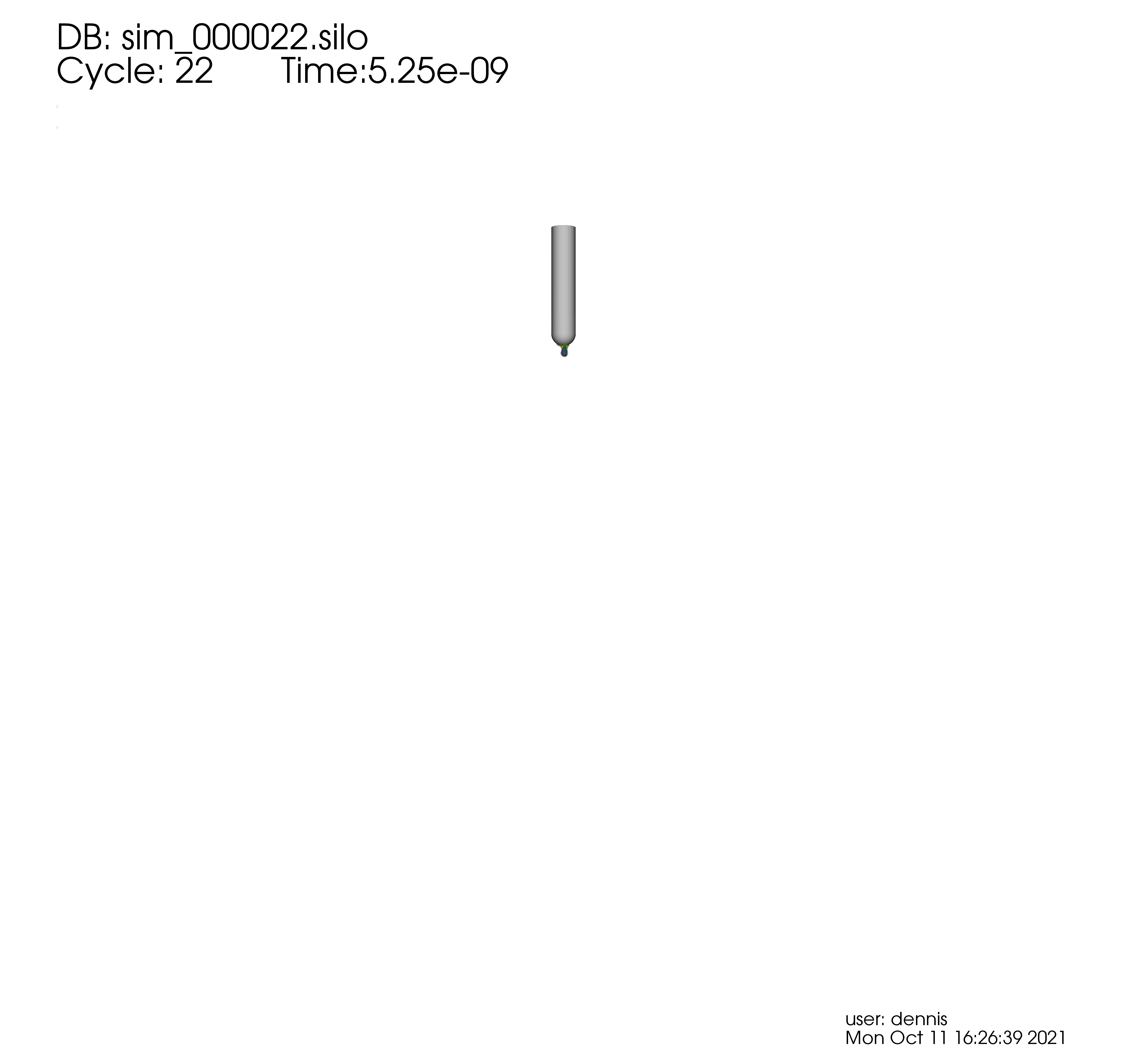}}
    \subcaption{$5.00$\,ns}
\end{minipage}
\begin{minipage}[t]{0.15\textwidth}
    {\includegraphics[trim=1600 1400 1600 750, clip, width=\textwidth]{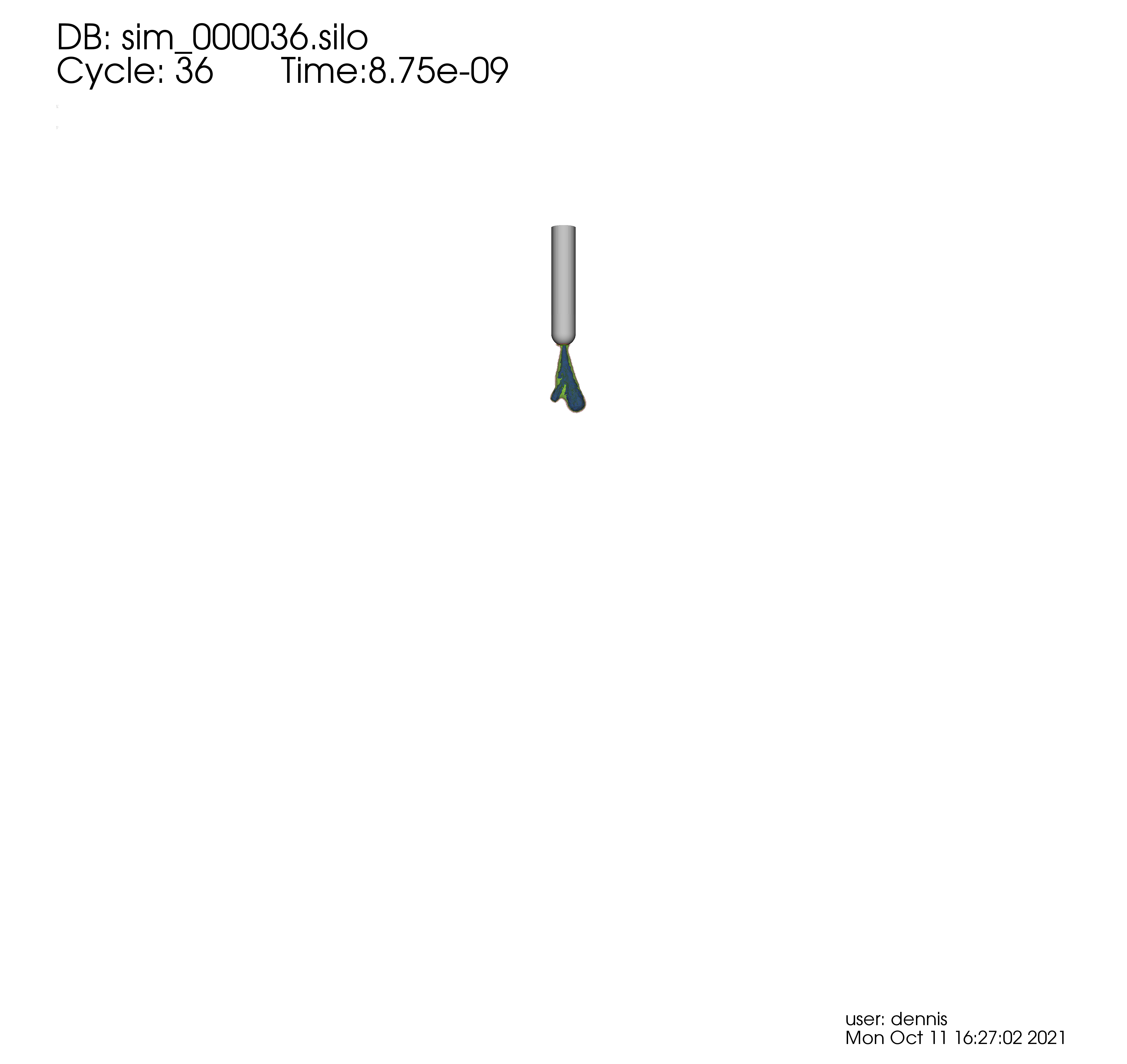}}
    \subcaption{$8.75$\,ns}
    \label{subfig:fuel_2}
\end{minipage}
\begin{minipage}[t]{0.15\textwidth}
    {\includegraphics[trim=1600 1400 1600 750, clip, width=\textwidth]{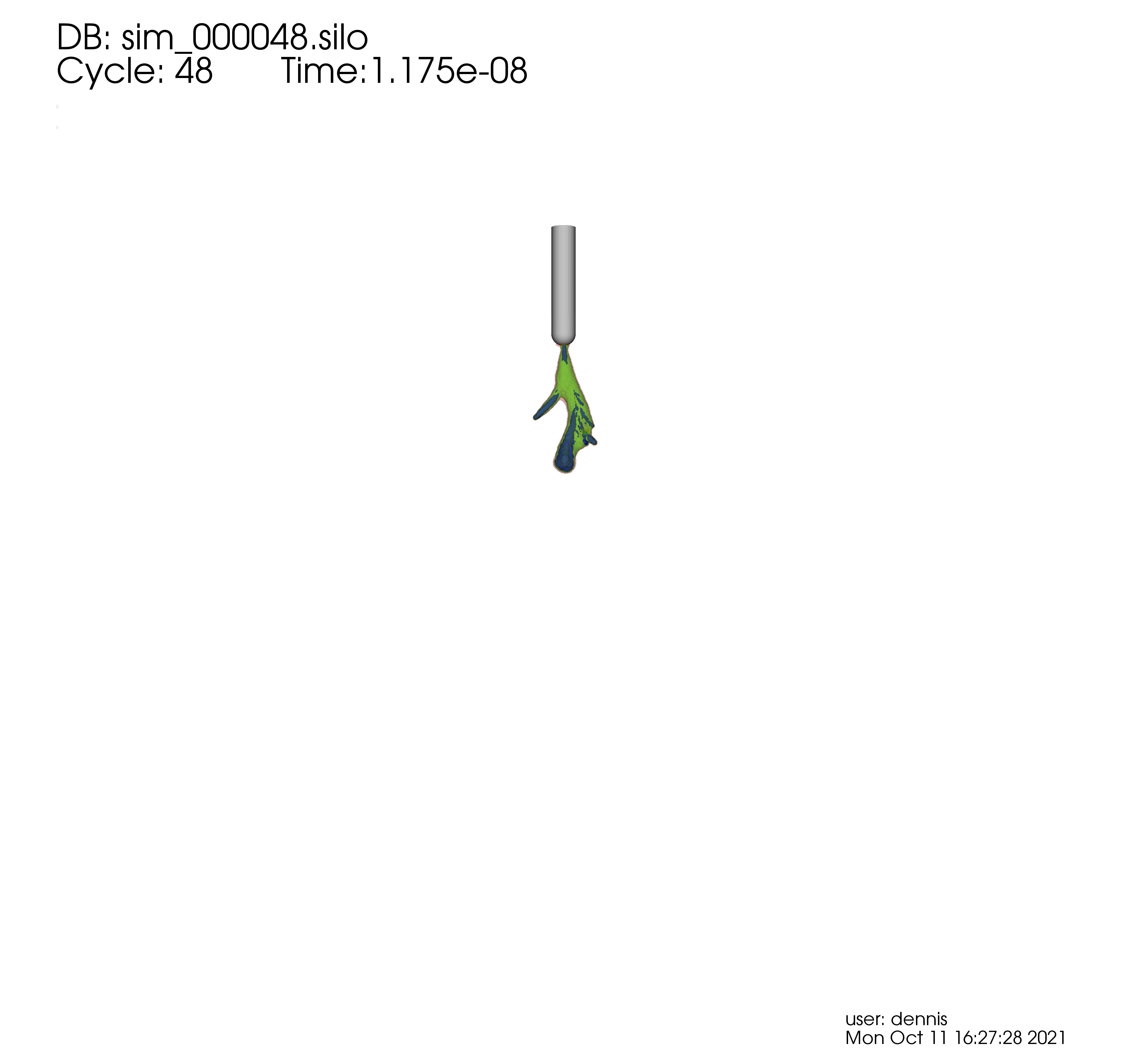}}
    \subcaption{$11.75$\,ns}
    \label{subfig:fuel_3}
\end{minipage}
\begin{minipage}[t]{0.15\textwidth}
    {\includegraphics[trim=1600 1400 1600 750, clip, width=\textwidth]{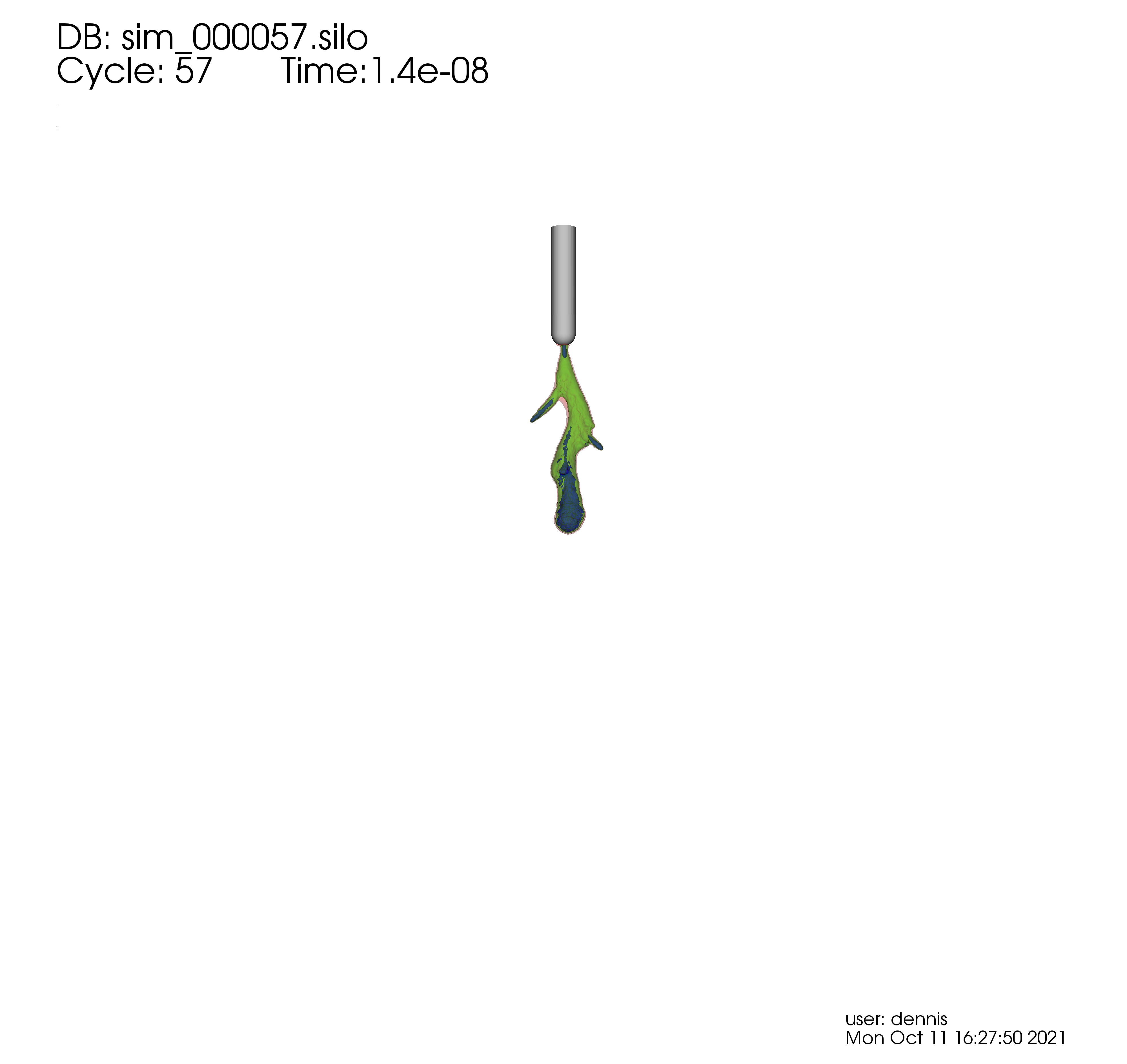}}
    \subcaption{$14.00$\,ns}
\end{minipage}
\begin{minipage}[t]{0.15\textwidth}
    {\includegraphics[trim=1600 1400 1600 750, clip, width=\textwidth]{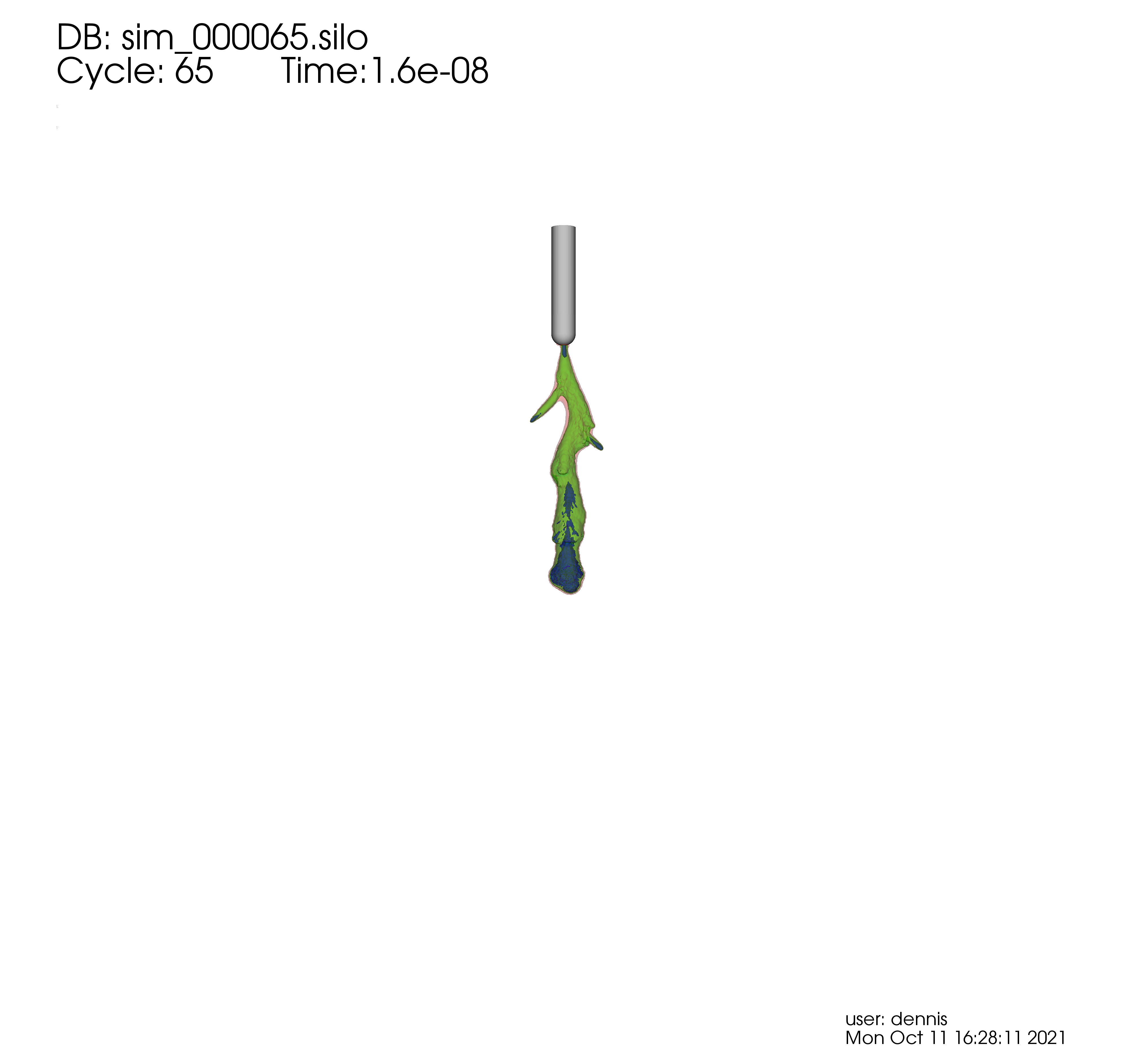}}
    \subcaption{$16.00$\,ns}
    \label{subfig:fuel_5}
\end{minipage}
\begin{minipage}[t]{0.15\textwidth}
    {\includegraphics[trim=1600 1400 1600 750, clip, width=\textwidth]{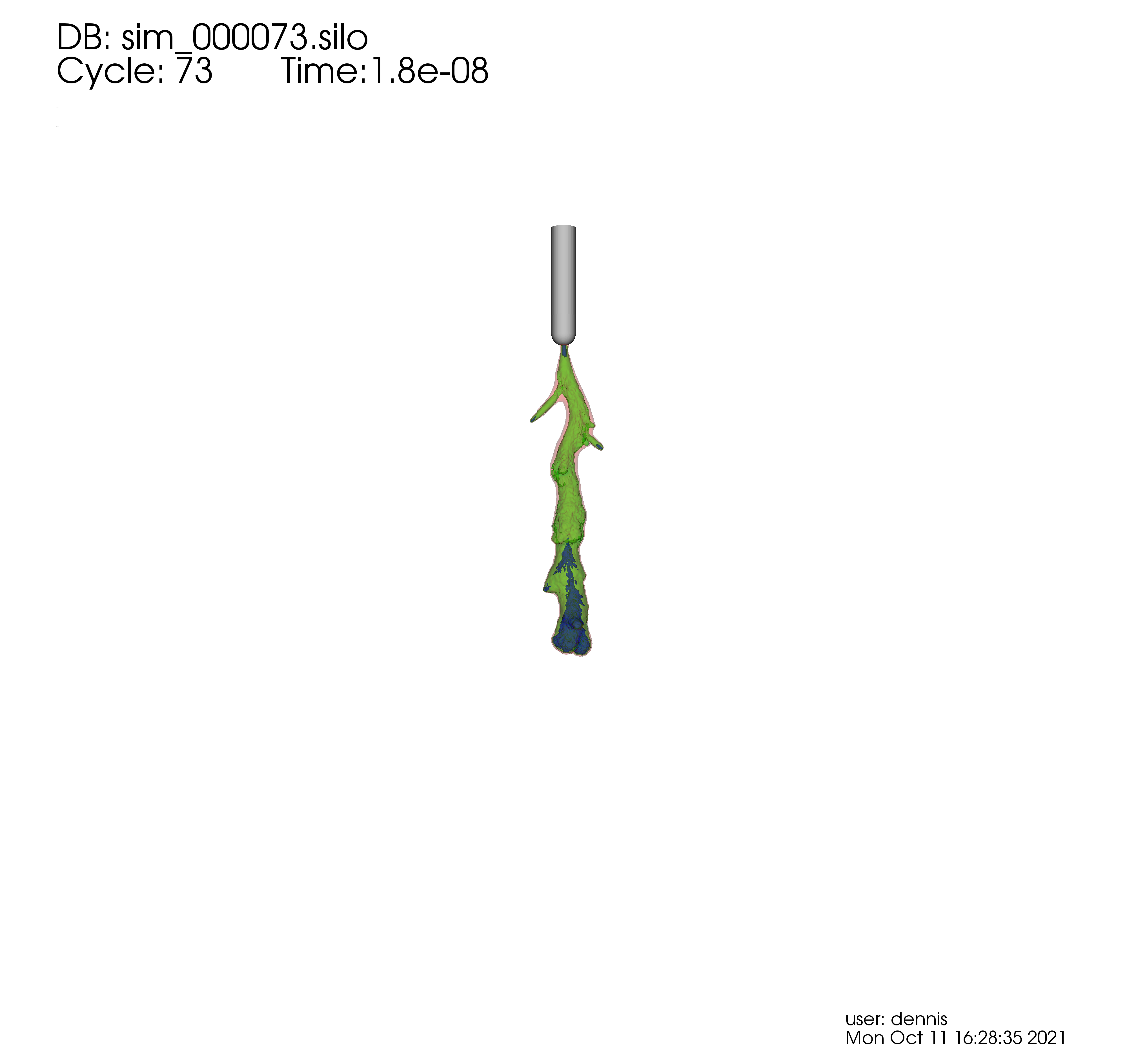}}
    \subcaption{$18.00$\,ns}
    \label{subfig:fuel_6}
\end{minipage}
    \caption{Evolution of the electron density for positive streamers as a function of streamer length. Streamers propagate in a background field of $E_0=12.5$\,kV/cm in air (top) and air-methane (bottom). The corresponding simulated times are supplied in the sub-caption. Visualization is performed with three partially-transparent contour surfaces. We observe that shape of the air streamer is smooth whereas the air-methane streamer is more erratic.}
    \label{fig:thicc_fuel_and_air}
\end{figure*}

We will now study the influence of methane on streamer branching. To that end we have performed simulations in air and a stoichiometric air-methane mixture under the conditions described in section \ref{sec:parameters}. The background electric field was taken as \mbox{$E_0=12.5$}\,kV/cm. Furthermore the comparison will be performed for streamers of equal (vertical) length $L_z$. $L_z$ is obtained by calculating the maximum vertical distance between the electrode and each point in the domain where the electron density exceeds $10^{19}$\,m$^{-3}$. We have chosen this threshold value because such densities are typically obtained in (or very near) the space-charge layer around the streamer tip. We visualize the streamer by plotting three partially-transparent contour surfaces.

In figure \ref{fig:thicc_fuel_and_air} we show the time evolution of positive streamers in both gases. We observe that the initial streamer formation takes about $2$\,ns longer in air-methane than in air (for \mbox{$L_z =0.1$\,m}), but that the main branch of both streamers then propagates at approximately the same instantaneous velocity ($0.47$\,mm/ns for both gasses at $5.1$\,mm). For a discharge in air no branching occurs. However, in these conditions we do observe that the streamer does not propagate in a straight line but in a meandering fashion. Its counterpart in air-methane is more irregular. The first branching event occurred about $3.75$\,ns after inception (figure \ref{subfig:fuel_2}). Further stochastic fluctuations occur throughout the evolution of the discharge which give the streamer an erratic shape. 

\begin{figure*}
\centering
\begin{minipage}[t]{.49\textwidth}
    \centering
    \includegraphics[width=\textwidth]{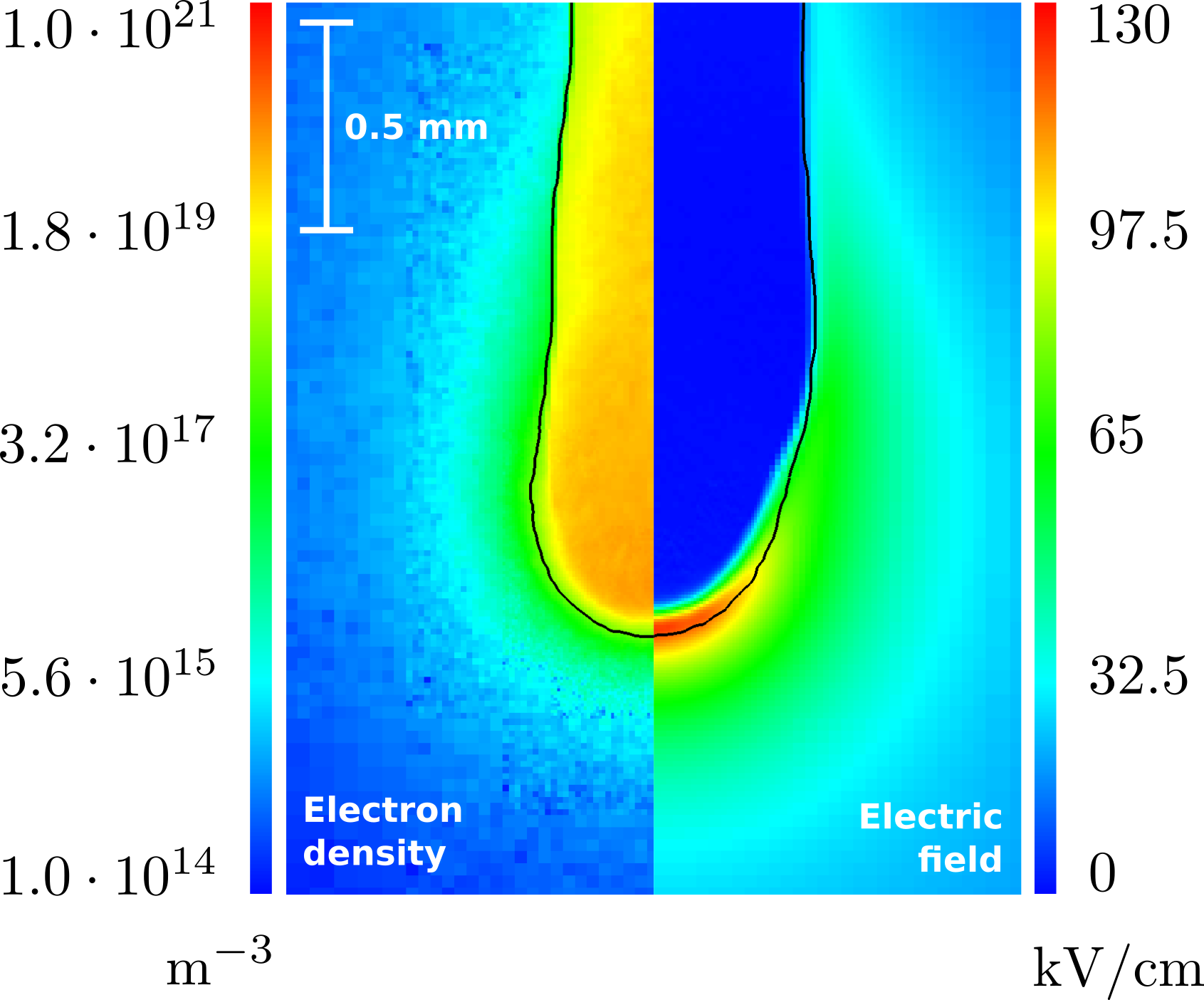}
    \subcaption{Air}
    \label{fig:test_leading_edge_air}
\end{minipage}\qquad
\begin{minipage}[t]{.49\textwidth}
    \centering
    \includegraphics[width=\textwidth]{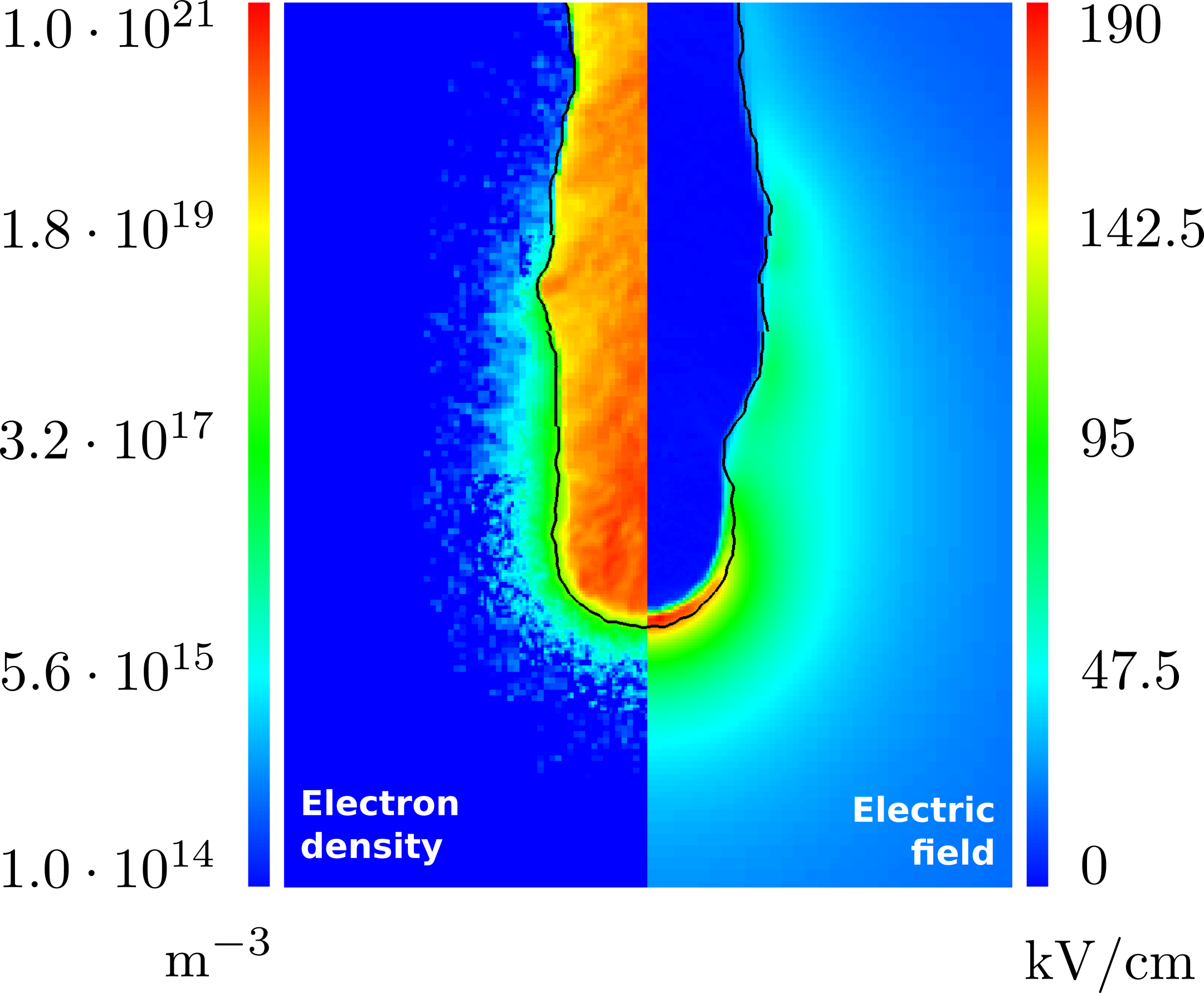}
    \subcaption{Air-methane}
    \label{fig:test_leading_edge_methane}
\end{minipage}
\caption{Zoom into the electron density and the electric field on a cross section through the streamer head, of the discharges shown in figure \ref{fig:thicc_fuel_and_air}. In all images, the contour line corresponding to an electron density of \mbox{$1.6\cdot10^{18}$}\,m$^{-3}$ is indicated. Electrons outside this contour are represented by individual particles (on the finest grid). Note the varying color scheme for the electric field which are matched to the electric field of the streamer tips. In air-methane we observe a smaller size of the electron cloud surrounding the streamer and a higher electric field at the tip than in air.}
\label{fig:leading_edge}
\end{figure*}

The erratic streamer shape due to the addition of methane can be attributed to the suppression of photoionization. In section \ref{sec:photoionization} we have shown that a gas fraction of $9.5\%$ methane already significantly reduced the rate of photoionization and the photon mean free path while only having a minor influence on the transport and reaction coefficients, as is also observed in \cite{Naidis2007}. This has a pronounced effect on the electron density ahead of the ionization front, as is shown in figure \ref{fig:leading_edge}. In this figure we see slices of the electron density and electric field along the axis of propagation for both streamers after $3.7$\,mm. Electron densities are shown on the same logarithmic color scale, with maximum densities in the channel of $7.5\cdot 10^{19}$\,m$^{-3}$ for air and $25\cdot 10^{19}$\,m$^{-3}$ for air-methane. Electric fields are shown on different linear color scales, which are matched to the electric field at the streamer tip ($130$\,kV/cm in air and $190$\,kV/cm in air-methane). In both gases we observe that the electron density in the leading edge decreases as we move farther away from the ionization front. However, for a streamer in air the electron density extends further ahead of the ionization front than its counterpart in air-methane. This is due to the difference in the longest photon mean free path. For sufficiently large distances $r$  we find that the asymptotic behaviour of the function $\Psi$ satisfies:
\begin{equation}
    \Psi(r)\propto r^{-1}e^{-r/l_\text{max}},
\end{equation}
with $l_\text{max}$ the longest photon mean free path (cf.\ figure \ref{fig:photon_meanfreepath} at $12.1$\,eV). Concretely, the characteristic length of the leading edge electron density, which is dominated by photoionization away from the streamer tip, is thus determined only by the longest photoionization length. Adding methane shortens the leading edge. Then, the electron density fluctuations ahead of the streamer tip, which are due to the discrete nature of photons, evolve due to impact ionization in the high electric field near the streamer tip. When these stochastically distributed electron avalanches reach the ionization front they accelerate branching \cite{Luque2011ElectronAir}. 

In section \ref{sec:particle_weight} we have mentioned how to mitigate the influence of artificial noise introduced by super-particles. We also argued that on the finest grid we simulate leading-edge electron densities up to \mbox{$1.6\cdot10^{18}$}\,m$^{-3}$ with single particles only. This threshold density is denoted in figure \ref{fig:leading_edge} by a black contour line. Since this contour line is close to the space-charge layer we conclude that the leading edge dynamics are properly resolved. Thus, the observed stochastic fluctuations and their influence on streamer branching are physical.

\subsection{Volume distributions of $E$ and $n_e$}\label{sec:ne_and_E}

\begin{figure}
    \centering
    \includegraphics[width=0.475\textwidth]{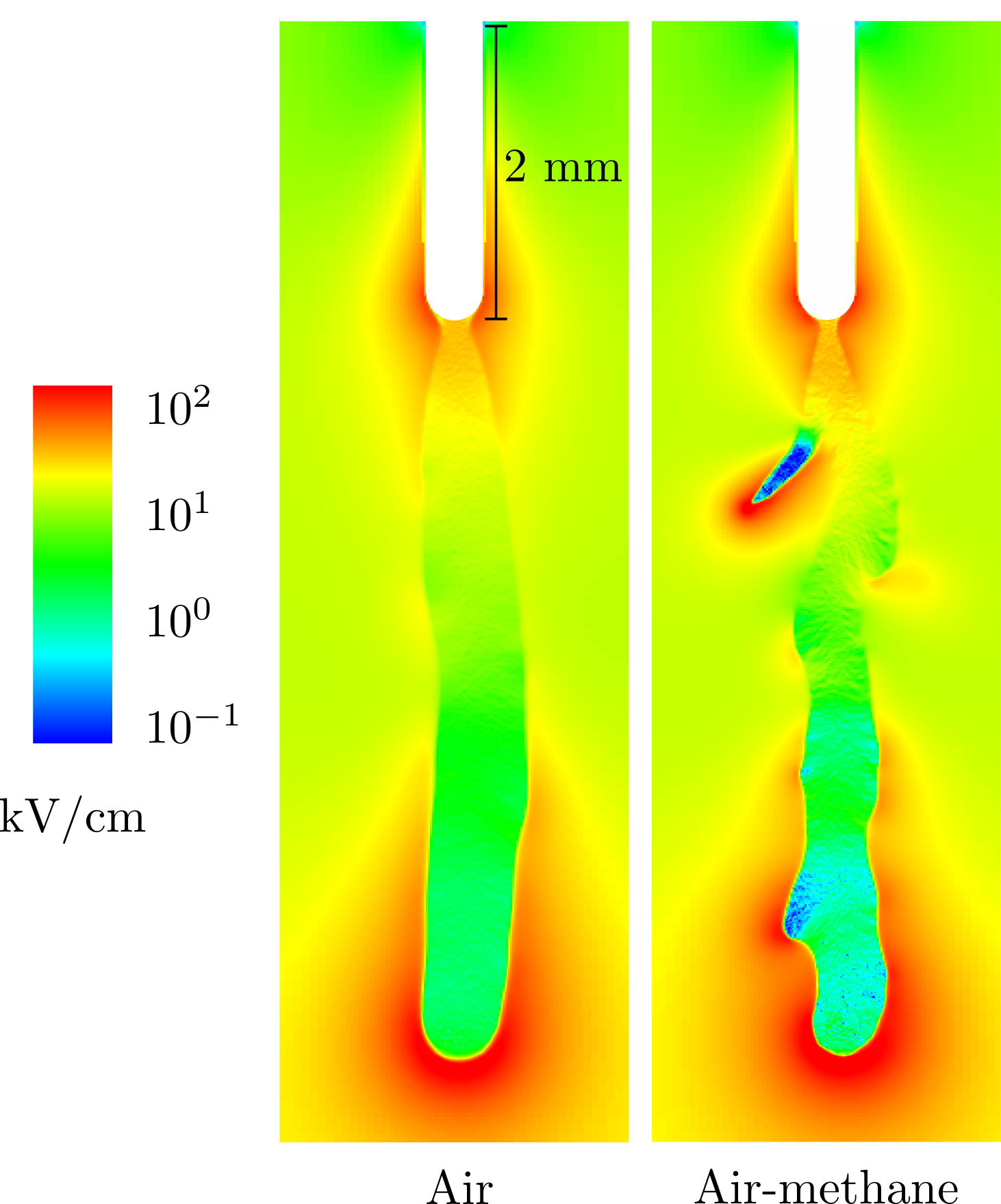}
    \caption{Cross sections of the electric fields of discharges shown in figures \ref{subfig:air_6} and \ref{subfig:fuel_6} on a logarithmic scale. Internal electric fields in the streamer are lower in air-methane than in air. The thin stagnated side-branch has the lowest internal electric field.}
    \label{fig:E_slices}
\end{figure}

We will analyze the quantitative difference in the electric field and electron density between the air and air-methane streamers presented in the previous section. From figure \ref{fig:thicc_fuel_and_air} one can already note that the electron density in the tips of the air-methane streamers is higher. Furthermore, figure \ref{fig:leading_edge} shows that electric field enhancement is higher at the main branch of the air-methane streamer. A more complete illustration of the electric fields is shown in figure \ref{fig:E_slices}. There we have shown cross sections of electric fields on a logarithmic scale for streamers in both gases at $5.1$\,mm. These cross sections have the same perspective as in figure \ref{fig:thicc_fuel_and_air} and slice through the middle of the electrode. Because the streamers are not perfectly axisymmetric,  sometimes streamer branches or other parts of the channel fall outside of the plane. Nevertheless, these images convey typical behaviour of the electric fields inside the streamer channel. In air, we observe that the electric field just after the streamer tip is around $1.5$\,kV/cm and gradually increases to around $25$\,kV/cm at the point of connection with the electrode. In air-methane typical electric fields inside the channel are slightly lower, in the range between $0.75-25$\,kV/cm. Exceptions are the thin stagnated side-branches in air-methane, where the smallest internal electric fields lie around $0.1$\,kV/cm.

\begin{figure*}
\centering
\begin{minipage}[t]{.49\textwidth}
    \centering
    \includegraphics[width=\textwidth]{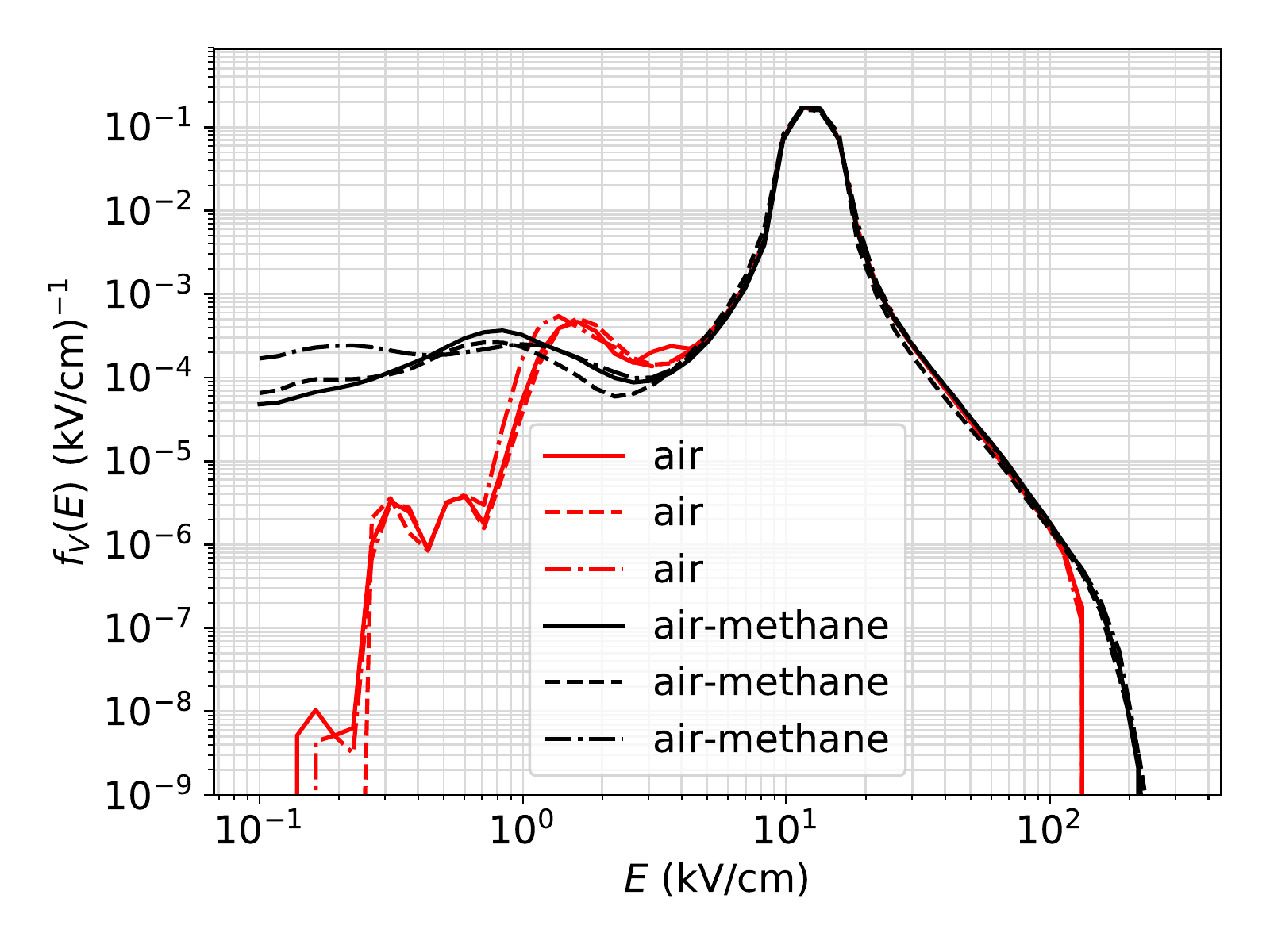}
    \subcaption{Electric field}
    \label{fig:Vdist_electric_field}
\end{minipage}\qquad
\begin{minipage}[t]{.49\textwidth}
    \centering
    \includegraphics[width=\textwidth]{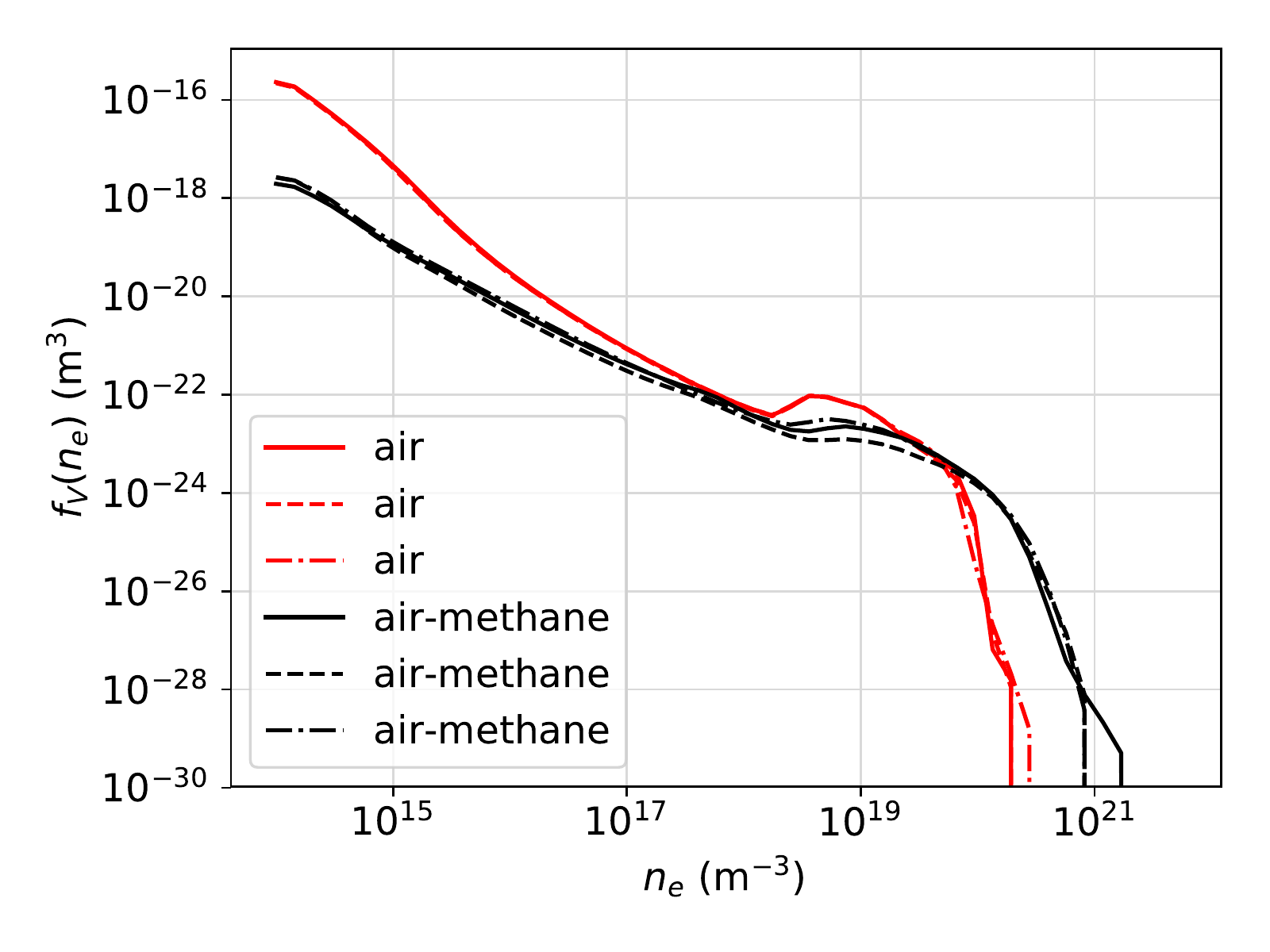}
    \subcaption{Electron density}
    \label{fig:Vdist_electron_density}
\end{minipage}
\caption{The volume distribution $f_V(E)$ of the electric field strength and $f_V(n_e)$ of the electron density. We have shown three air streamers and three air-methane streamers at a length of $5.1$\,mm under conditions corresponding to figure \ref{fig:thicc_fuel_and_air}. Air-methane streamers exhibit stronger electric field enhancement and higher electron densities.}
\end{figure*}

We will make these observations quantitative by introducing the volume-normalized distribution of the electric field strength $f_V(E)$:
\begin{equation}
    f_V(E) = \frac{1}{V_\mathrm{total}} \frac{V(E) - V(E + \Delta E)}{\Delta E},
\end{equation}
where  $V(E)$ is the volume in which the electric field strength exceeds $E$, and $V_\mathrm{total}$ is the total volume. So $f_V(E)$ denotes the volume of the domain where the electric field strength lies within the small interval within $E$ and $E+\Delta E$ which is then divided by $\Delta E$ and normalized with respect to the total volume.

$f_V(E)$ is shown in figure \ref{fig:Vdist_electric_field} for three simulations per gas mixture with different random seeds for streamers of $5.1$\,mm length.  We use log-spaced bin sizes $\Delta E$ (for computational efficiency). The region with values below $5$\,kV/cm, corresponds to internal electric fields as shown in figure \ref{fig:E_slices}. The very low fields $0.1-0.5$\,kV/cm are due to small side branches. Since the streamers in air that we consider have not branched these regions are non-existent (up to stochastic fluctuations).  In the main channel of the discharge we find values of $1.5$\,kV/cm in air and $0.75$\,kV/cm in air-methane (in figure \ref{fig:E_slices} it is shown that internal electric fields gradually increase to $25$\,kV/cm near the electrode). This is reflected in the distribution function by peaks at the associated values. Moving to higher fields, we find a large volume of the domain with fields ranging from $10-15$\,kV/cm. This interval is dominated by the region away from the streamer, where the background electric field persists. Moving to even higher fields, we find the region that corresponds to the active zone induced by the discharge. Here we observe that air-methane streamers exhibit higher values for the electric field in its active zone. The maximum electric field is around $215$\,kV/cm in air-methane whereas in air we find a maximum value around $120$\,kV/cm. The maximal field in air-methane corresponds to the thin stagnated side-branch on the left (see figure \ref{subfig:fuel_6}).

Analogously we show the volume distribution of the electron density:
\begin{equation}\label{eq:distribution_function}
    f_V(n_e) = \frac{1}{V_\mathrm{total}} \frac{V(n_e) - V(n_e + \Delta n_e)}{\Delta n_e}
\end{equation}
in figure \ref{fig:Vdist_electron_density}. 
Here  $V(n_e)$ is the volume in which the electron density exceeds $n_e$.
Air-methane streamers typically exhibit higher electron densities, which is probably due to a higher electric field at the tip. Furthermore, the low electron density region, say below $10^{16}$\,m$^{-3}$, corresponds to the electron `cloud' surrounding a streamer which is a result of photoionization. We observe that in air this electron cloud fills a much larger volume than in air-methane. In air-methane, the size of the cloud is reduced because \CH{4}{} shortens the photon mean free path. Thus we find that a smaller part of the discharge is associated with this region.

To conclude, positive streamers in air-methane compared to air: (1) have smaller electron cloud surrounding the streamer, (2) have electric fields at the tips of their ionization fronts which are higher by a factor $1.5$, (3) have internal electric fields that are lower by a factor two and (4) have higher electron densities that are higher by a factor three.

\subsection{Electron energy distribution}

\begin{figure}
    \centering
    \includegraphics[width=0.49\textwidth]{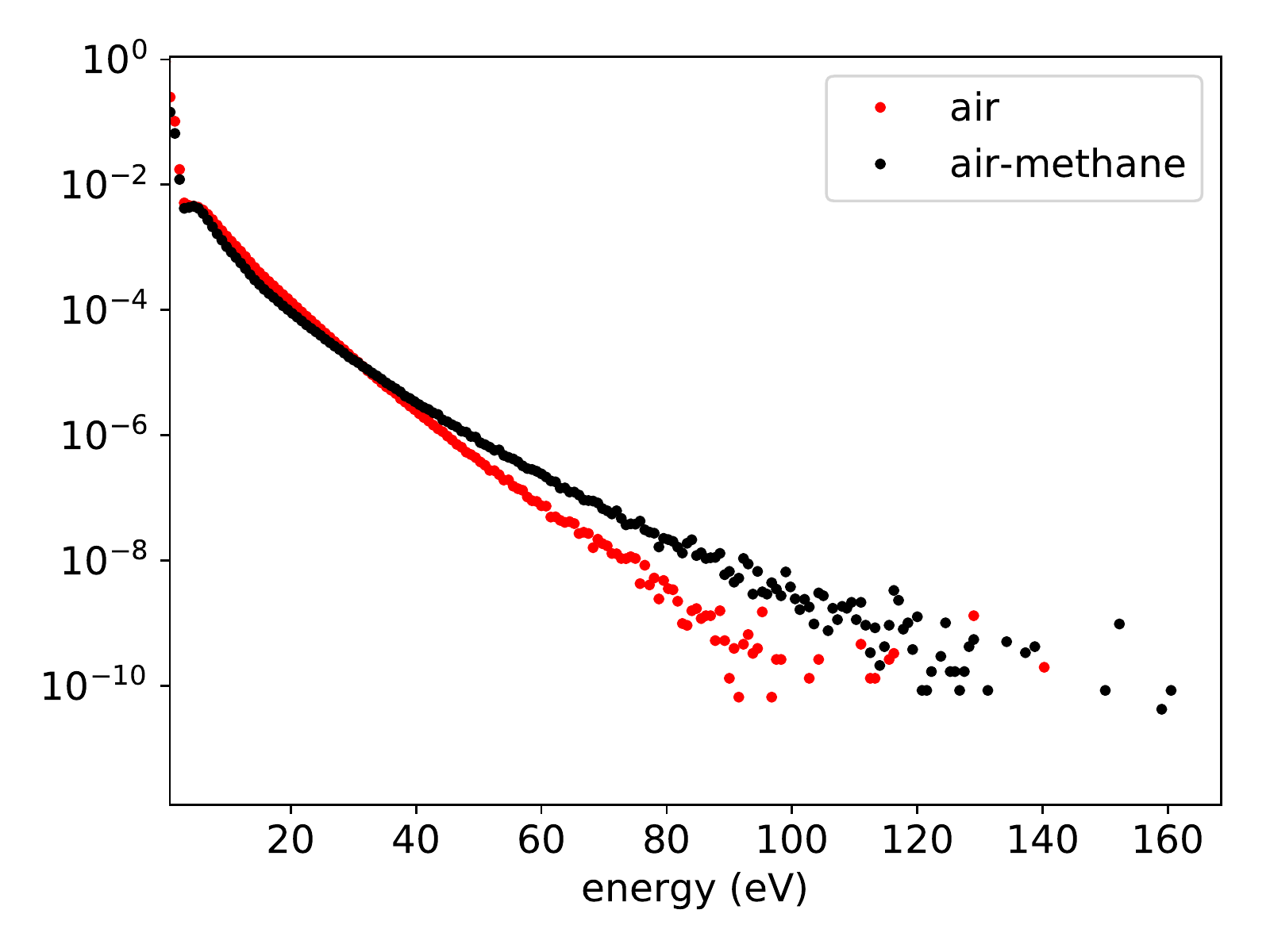}
    \caption{The EEDF of streamers presented in section \ref{sec:branching} with a length of $5.1$\,mm. A higher $E_{\text{max}}$ in the case of air-methane leads to higher electron acceleration which enhances the tail.}
    \label{fig:EEDF}
\end{figure}

In figure \ref{fig:EEDF} we show a comparison of the electron energy distribution function (EEDF) of positive streamers in air and a stoichiometric air-methane mixture. The distribution was calculated in both gases when the streamer reached a length of $5.1$\,mm. This corresponds to figures \ref{subfig:fuel_6} and \ref{subfig:air_6}. The EEDF was obtained by calculating the kinetic energy of each super-particle and making a histogram with a bin size of $0.75$\,eV. 

For both gases we observe similar qualitative behaviour of the EEDF: most of the electrons have energies below $5$\,eV and only a few electrons have energies sufficiently high to ionize neutral gas molecules (e.g.\ the lowest ionization threshold is $12.1$\,eV corresponding to \O{2}{}). These electrons are most likely to be situated in the region where ionization occurs, namely the ionization front and the part of the channel close to the electrode. For both gases we also find electrons with energies above $100$\,eV. For air-methane we even observe an electron energy exceeding $160$\,eV, which can be considered high for positive streamers. Electrons with such high energies are approaching the cold-electron runaway regime \cite{Diniz2018ColdCurve} and challenge the assumption of isotropic scattering, which is also made in our model.\\

A distinction between the two gases is that the tail of the distribution is more enhanced in the air-methane mixture. This is to be expected due to the higher electric field at the tip of the streamer, as was shown in section \ref{sec:ne_and_E}, combined with the fact that electron energy losses in the tail hardly change by the addition of methane~\cite{Kohn2019}, cf.\ figure \ref{fig:swarm_parameters}.

\section{Plasma-chemical activation} \label{sec:kinetics}

Here we will investigate the energy deposition and the production of reactive species. These streamer properties are relevant for plasma-assisted combustion. Furthermore we highlight that these properties are comparatively insensitive to the considered electric fields.

\subsection{Deposited energy density}\label{sec:energy_transfer}

\begin{figure*}[!htb]
\centering
\includegraphics[width=\textwidth]{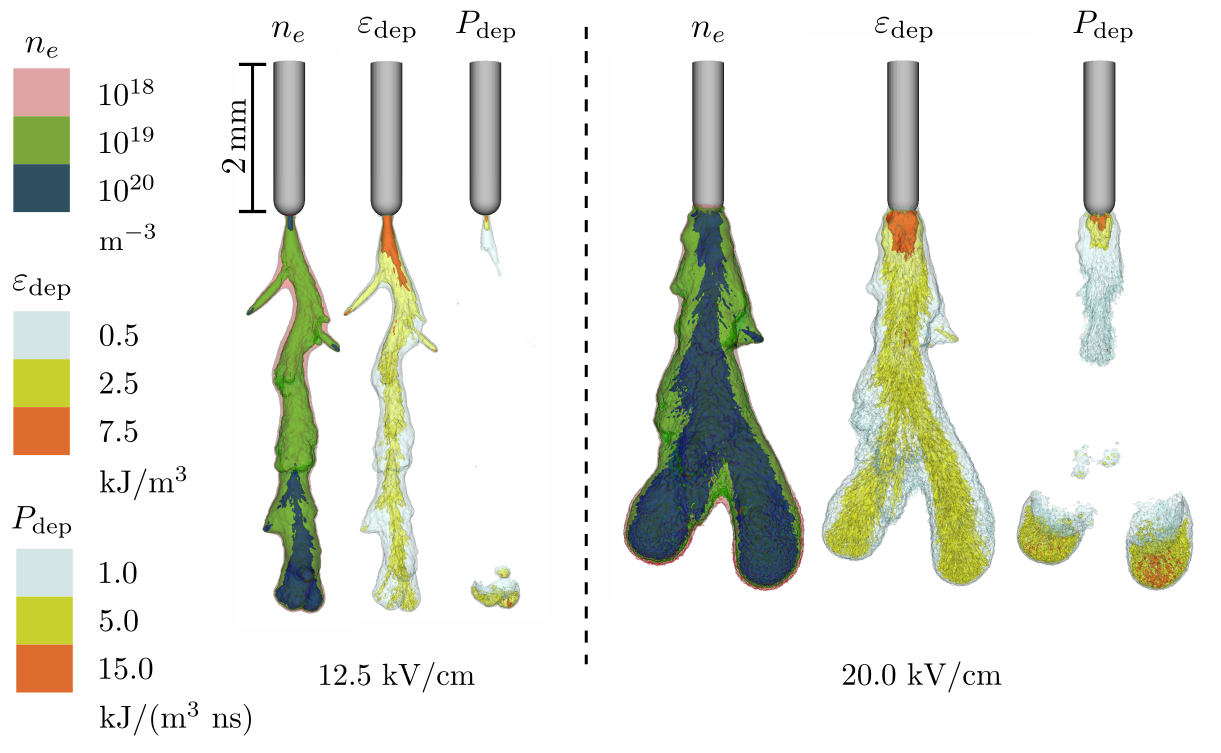}
\caption{Contour surfaces of the electron density, the deposited energy density and the power density deposition within $0.25$\,ns. We show air-methane streamers with a length of \mbox{$L_z=4.8$}\,mm in two background fields. Due to a higher velocity in higher electric fields, the contribution of the ionization front to $P_{\rm dep}$ appears more smeared out.}
\label{fig:E_dep}
\end{figure*}

We will now investigate the energy that a positive streamer deposits to the gas molecules in a stoichiometric air-methane mixture. To this end, we have performed simulations under the conditions shown in figure \ref{fig:simulation_domain} for two background fields: \mbox{$E_0=12.5$} and $20$\,kV/cm (named `low field' and `high field', respectively). From these simulations we have extracted the deposited energy density by electron scattering $\varepsilon_{\rm dep}$.  Note that this quantity is integrated over time. The deposited energy density is calculated by cumulatively interpolating the energy losses of all the inelastic scattering events for each time step to the grid. Since inelastic electron scattering is the dominant contribution to the deposited energy density, this term represents the conversion of kinetic electron energy to chemical activation. In a fluid approach, the deposited energy density described above corresponds to the term \mbox{$\int_0^T j\cdot E\,dt $}, where $j$ is the electric current density and $T$ is the time. Similarly we also calculate the power density deposition $P_{\rm dep}$, which is the instantaneous energy density deposition. This quantity is obtained by calculating the numerical time derivative of $\varepsilon_{\rm dep}$ with a time step of $0.25$\,ns. Since the high field streamers have a higher velocity, the contribution corresponding to the moving ionization front appears to be more smeared out compared to the low field.

In figure \ref{fig:E_dep} we have shown the contour surfaces of the electron density $n_e$, the time-integrated deposited energy density $\varepsilon_{\rm dep}$ and the power density deposition $P_{\rm dep}$ for both applied electric fields. Both streamers have an equal length of \mbox{$L_z=4.8$}\,mm. In the low field the discharge developed to this size in $18$\,ns whereas in the high field only $4.25$\,ns was needed. A result of the difference in the timescales is that the effects of electron loss in the channel due to attachment are visible in the low field but not in the high field. For the $\varepsilon_{\rm dep}$ we observe a peak in deposited energy in a small region close to the electrode in both fields. In that region the energy density ranges from $10$ to $100$\,kJ/m$^{3}$ (not shown) and keeps growing over time. Such interactions between the streamer channel and the needle electrode are not uncommon, see \cite{Komuro2012, Li2021ComparingValidation} for example.   
Away from the needle we have a typical deposited energy density between $0.5$ and $2.5$\,kJ/m$^{3}$ that is comparatively insensitive to the applied electric field. However, for thin branches (as one can see on the sides of the discharge in the low field) the deposited energy is higher: around \mbox{$5.0$\,kJ/m$^{3}$}. These thin side-branches have a higher deposited energy density than the thicker main streamers due to a higher electric field enhancement at the streamer tip. For the contour surfaces of $P_{\rm dep}$, we have chosen the contour values such that they correspond to the deposited energy density within the time step of $0.25$\,ns. From this we can see that energy is mainly deposited by the ionization front and the small region near the electrode. As the ionization front propagates, it leaves behind energy in the new region of the channel. Moreover, as the streamer grows further a current continuously flows through the channel which also contributes to the deposited energy density, but these illustrations indicate that this contribution is only minor. Furthermore, the large power density deposition near the electrode shows that the deposited energy in this region grows over time.

\begin{figure}
    \centering
    \includegraphics[width=0.5\textwidth]{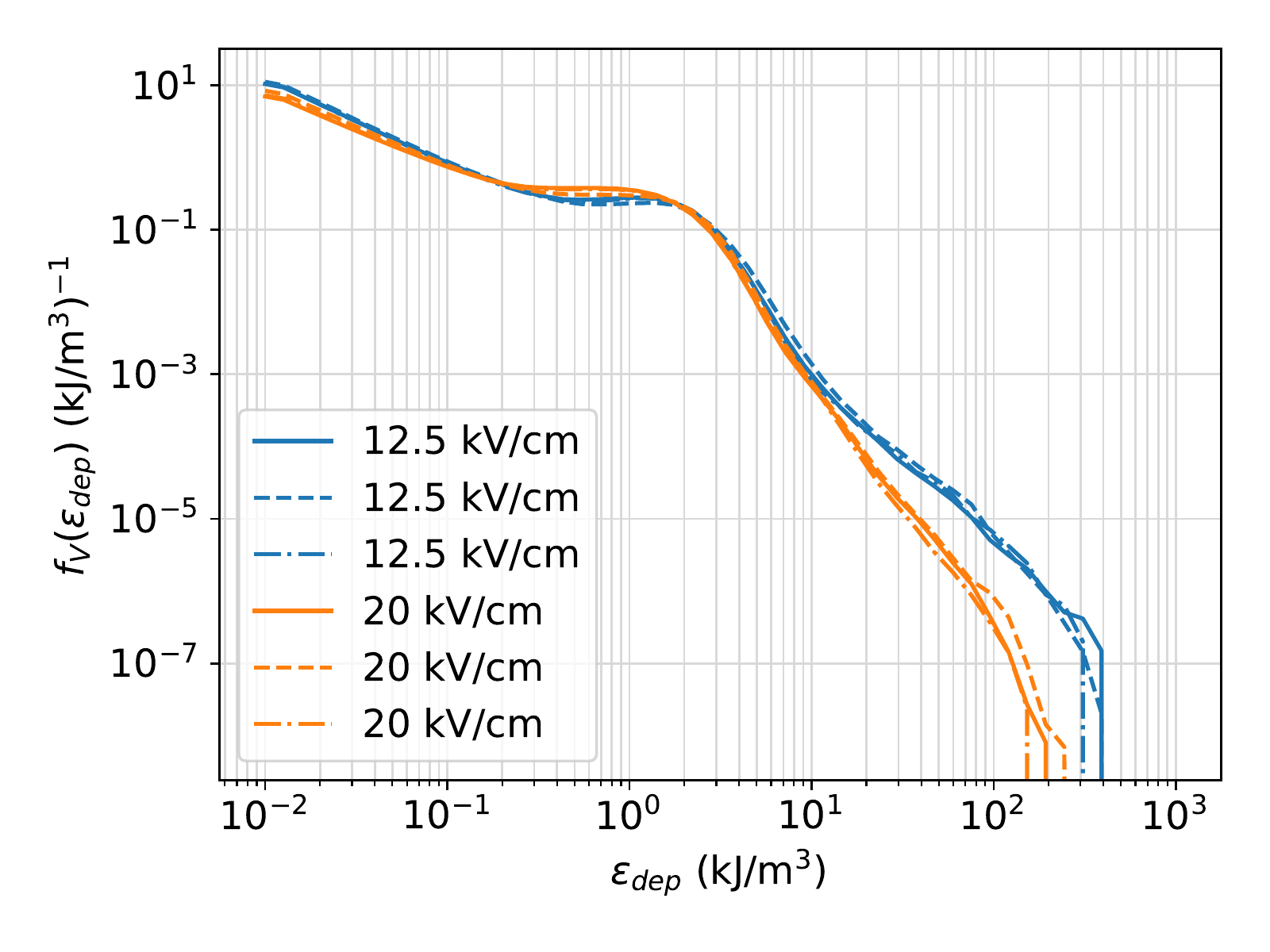}
    \caption{Volume distribution of the deposited energy density for air-methane streamers with \mbox{$L_z=4.8$\,mm}. For each applied electric field we have performed three simulations.}
    \label{fig:energy_Vdist}
\end{figure}

In figure \ref{fig:energy_Vdist} the normalized volume distribution of the deposited energy density $f_V(\varepsilon_{\rm dep})$ is shown for three simulations per applied electric field. This quantity is defined analogously to equation \eqref{eq:distribution_function} but with \mbox{$V_{\rm total} = V(\varepsilon_{\rm dep} > 10^{-2}$\,kJ/m$^3$}). In other words we normalize with respect to the volume treated by the discharge, which is taken as the volume where the deposited energy density exceeds $10^{-2}$\,kJ/m$^3$. This leaves the volume distribution invariant to differences in amount of volume treated by the discharge. This is convenient when comparing discharges that have different radii as is the case for streamers in different electric fields. From this graph we can conclude that a higher applied electric field does not necessarily lead to a higher deposited energy density, as the distributions of both fields (up to $10$\,kJ/m$^{3}$) practically coincide. Note that the high field discharge does in fact treat a larger volume of the gas, but that distribution functions are invariant to such differences due to normalization. Since the deposited energy density below $10$\,kJ/m$^{3}$ is associated with the ionization front, cf.\ the contour surfaces of figure \ref{fig:E_dep}, These results indicate that the deposited energy density dynamics of the ionization front are quite insensitive to the applied electric fields considered here.

Furthermore, for deposited energy density values above $10$\,kJ/m$^{3}$ (which are associated to the region close to the needle electrode) the distribution is actually higher in the low electric field. One aspect that contributes to this difference is  that the width of the channel that is connected to the electrode is much smaller for the streamer in the low field. Hence the current density that flows through the channel of the streamer is higher near the electrode which leads to a larger deposited energy density.

In conclusion: Away from the electrode, streamers in both fields considered here have a similar deposited energy density. However, the high field streamers are generally wider and therefore treat a larger volume of the gas.

\subsection{G-values for the production of reactive species}\label{sec:CAS}

A positive streamer produces various reactive species that are of interest to plasma-assisted combustion. For example, electron-impact can directly dissociate a molecule and hence produce radicals. Moreover, the electronically excited states can also lead to further radical production by auto-dissociation, dissociative quenching of another molecule or through chemical reactions \cite{Starikovskiy2012}. \\

In this section we will calculate the efficiency with which specific primary species are generated. With `primary' we mean that we only consider the production that follows directly from electron impact due to the streamer. The efficiency is expressed as the number of reactive particles created per $100$\,eV of input energy, named the G-value (with units \mbox{1/(100 eV)}). In order to calculate this we have cumulatively recorded the number of excitations that have occurred for each type of collision. The spatial coordinates of an excitation event are not stored, as that would significantly affect computation time.  \\

The G-values corresponding to individual species are shown in \ref{app:G-values}. 
In this section, the species are grouped into:
\begin{itemize}
    \item the oxygen singlet states \O{2}{}$(a^1\Delta_g$, $b^1\Sigma_g^+)$, 
    \item the nitrogen triplet states \N{2}{}$(A^3\Sigma_u^+$, $B^3\Pi_g$, $C^3\Pi_u$, $B'^3\Sigma_u^-$, $E^3\Sigma_g^+$, $F^3\Pi_u$, $G^3\Pi_u$, $W^3\Delta_u)$, and %\N{2}{}$(A^3\Sigma_u^+)$,  \N{2}{}$(B^3\Pi_g)$, \N{2}{}$(C^3\Pi_u)$, \N{2}{}$(B'^3\Sigma_u^-)$, \N{2}{}$(E^3\Sigma_g^+)$, \N{2}{}$(F^3\Pi_u)$, \N{2}{}$(G^3\Pi_u)$ and \N{2}{}$(W^3\Delta_u)$, and 
    \item the radicals N, O, H and \H{2}{} that are directly created by electron impact dissociation.
\end{itemize}
The spatial profile of the densities of all species (not shown) is similar to that of the energy deposition depicted in section \ref{sec:energy_transfer}. In the region near the needle-electrode we will find high values of the densities. Away from the needle the reactive species are produced in comparatively fixed densities.\\

\begin{figure}
    \centering
    \includegraphics[width=0.5\textwidth]{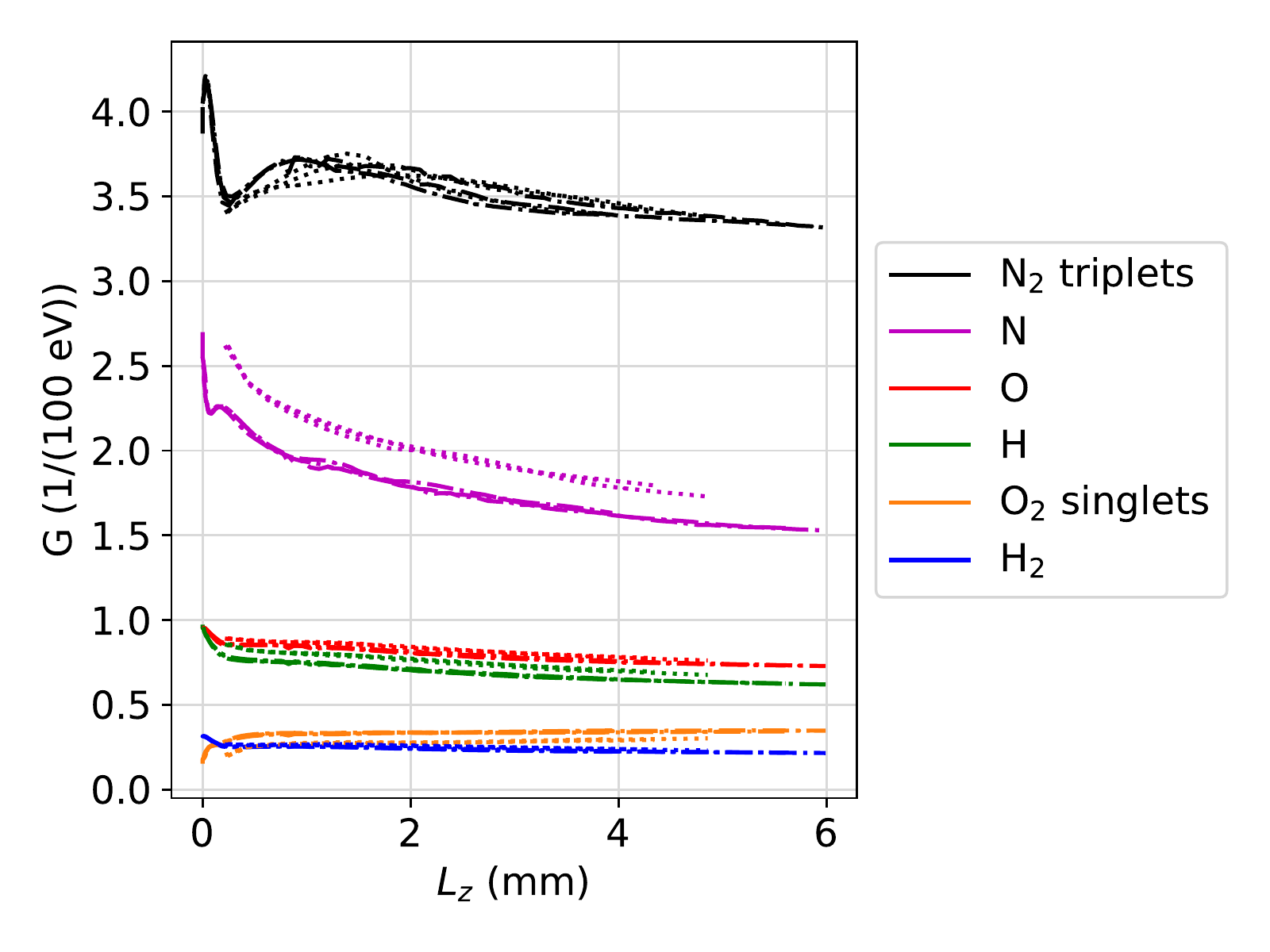}
    \caption{The G-values (i.e.\ number of particles produced per $100$\,eV) of various groups of reactive species as a function of streamer length $L_z$. Dash-dotted corresponds to the low background field of $12.5$\,kV/cm. Dotted corresponds to the high background field of $20.0$\,kV/cm.}
    \label{fig:G_values}
\end{figure}

In figure \ref{fig:G_values} we show the G-values produced by the discharges presented in section \ref{sec:energy_transfer}, for the background fields ${E_0=12.5}$ and $20.0$\,kV/cm. The G-values are calculated for the whole discharge evolution until the streamer has reached the vertical length $L_z$, and they are plotted as a function of $L_z$. Similar to the results of the average energy deposition, we find that the G-values for the production of reactive species are comparatively insensitive to the applied electric field and do not change greatly as the streamers grow. Similar observations were made regarding the calculated G-values in flue-gases \cite{Babaeva1998Two-DimensionalGaps}. The highest G-value corresponds to nitrogen triplet states, where $3.3$ particles are produced per $100$\,eV (at a length of $5$\,mm). Direct electron impact dissociation of molecules nitrogen and oxygen is also prevalent, with G-values of \mbox{${1.5-1.7}$} and ${0.7}$, respectively. Furthermore, under these conditions the electron-impact dissociation of methane produces atomic and molecular hydrogen radicals with a typical G-value of ${0.6}$ and ${0.2}$, respectively. Finally, the low-threshold singlet states of oxygen have a comparatively low G-value of around ${0.3}$.\\

These G-values show that positive streamers in stoichiometric air-methane mixtures primarily produce large amounts of reactive nitrogen species. This leads to further dissociation of \O{2}{}, \CH{4}{} and \H{2}{} in the plasma-afterglow (which is not accounted for here). For \O{2}{}, we observe that direct dissociation by electron impact is much more prevalent than excitation to the low-lying singlet states. Finally, the dissociation of \CH{4}{} predominantly produces atomic hydrogen, in amounts which are approximately equal to atomic oxygen.\\

In plasma-assisted combustion there is an interest in a broad range of operating conditions, with notable emphasis on high pressures and temperatures. In order to address that interest we  sketch how to generalize our results to a wider parameter regime. We do so by relying on scaling laws for electric discharges, as is explained in \cite{Nijdam2020ThePhenomena}. Discharges at different pressures and temperatures are physically similar if the reduced electric field $E/N$ is the same. Note then that high pressure and temperature can partially compensate each other as the relevant quantity in discharge physics is the gas density $N$. The G-values reported here are quite insensitive to the (reduced) electric field, suggesting that they might be representative for a larger range of pressures and temperatures than was considered in this work. Having said all that, it is known that the applicability of scaling laws is limited by a number of effects. For instance, at standard temperature and pressure the three-body attachment of electrons and the suppression of photon emission by collisional quenching violate these scaling laws. Thus electron loss and the production of photoelectrons do not scale with $E/N$. Another limiting factor is that we assume that electrons collide only with ground-state molecules. However, for very high temperatures or partially burnt mixtures electron scattering with fragmented or excited species will no longer be negligible.\\

Our results are relevant for studies using global (i.e.\ quasi-0D) kinetic models. These models are used to simulated the time evolution of hundreds of species using detailed chemical reaction mechanisms within reasonable computation time. However, such models can not resolve the streamer discharge phase which is strongly non-uniform, both spatially and temporally. The G-values reported here provide insight into the initial (practically instantaneous) production of reactive species. In this regard we mention other modelling studies that have reported on the production of reactive species in air-methane mixtures by a streamer discharge  \cite{Breden2013, Breden2019, Naidis2007, Takana2014, Levko2017, Qian2018FluidDischarge, Yang2017}. However, as far as the authors are aware the G-values for the primary production phase have not been computed with the level of detail and as complete as they have been reported here.

\section{Summary and Discussion}\label{sec:conclusion}

\subsection{Summary}
We have performed three-dimensional particle-in-cell simulations of positive streamers in air and in stoichiometric air-methane. In order to do so, we first extended the stochastic photoionization model by Zelezniak~\textit{et al.}~\cite{Zhelezniak1982PhotoionizationDischarge} to account for quenching and non-ionizing photoabsorption due to methane. It follows that $9.5\%$ methane already significantly suppresses photoionization in air, decreasing both number and mean free path of photons within the relevant energy interval significantly. Subsequently this influences streamer properties. This model allowed us to show that:
\begin{enumerate}
    \item air-methane streamers branch more often than their counterparts in air. We have attributed this to the decreased  photoionization, while mobility and effective ionization coefficients stay essentially unchanged.
    \item under the same conditions streamers in air-methane have a higher field enhancement than in air, lower internal electric fields, a higher electron density in the streamer channel, and higher electron energies.
\end{enumerate}
On the side of plasma-assisted combustion we have shown that:    
\begin{enumerate}[resume]
    \item the ionization front in air-methane streamers in background fields of $12.5$ and $20.0$\,kV/cm typically deposits around $0.5-2.5$\,kJ/m$^{3}$ into plasma-chemical activation of the gas. This quantity appeared to be quite insensitive to the considered applied electric fields.
    \item  similar to the deposited energy density, the streamers produce typical densities of reactive species within the streamers that appear to be quite insensitive to the electric fields applied in this study. We calculated the G-values for the production of these species which are primarily \N{2}{} triplet states, but also N, O and H radicals.
    \item aside from aforementioned similarities, the high field streamers have larger radii (thus they treat a larger volume) and propagate faster.
\end{enumerate}

\subsection{Discussion}
Our comparison of streamers in air and in a stoichiometric air-methane mixture shows that it is important to correct the photoionization properties for the presence of methane. However, the importance of a suppressed photoionization mechanism diminishes in situations with considerable degrees of background ionization, for instance in a pulsed discharge with a high repetition frequency.

On the side of plasma-assisted combustion, our model represents only the initial source of reactive species that arises due to direct electron impact or photoionization in the streamer discharge. Important secondary production of radicals due to chemical reactions occurs on slower time scales. The G-values presented in this work could be used in a combustion model adapted to resolve the subsequent chemical processes and thereby account for the non-equilibrium excitation of the gas by a streamer. In this context we have shown that, away from the needle-electrode, the streamer produces reactive species in comparatively fixed densities which suggest that a volume-averaged modelling approach with a parametrization of the streamer phase can be considered. We refer the reader to \cite{Levko2017} for a deeper discussion into the use of global models.

\section*{Acknowledgements}
This work was carried out on the Dutch national e-infrastructure with the support of SURF Cooperative. The three-dimensional renders presented in this work have been made using VisIt \cite{HPV:VisIt}. DB acknowledges funding through the Dutch TTW-project 16480 `Making Plasma-Assisted Combustion Efficient'.

\appendix

\section{Tabulated G-values per excited species}\label{app:G-values}
In section \ref{sec:CAS} and in figure~\ref{fig:G_values} the G-values of groups of excited species were given. Here we resolve the G-values per excited species in tables \ref{tab:G_N2}, \ref{tab:G_O2} and \ref{tab:G_CH4} for \N{2}{}, \O{2}{} and \CH{4}{}, respectively. The G-values are the number of excitation events per $100$\,eV that generate a particular excited species or group of species. Note that to convert back to the G-values per grouped species, the values corresponding to processes that produce multiple particles first have to be properly weighted, e.g. the reaction:
$$ \text{e + \N{2}{} $\to$ e + N + N},$$
produces two nitrogen atoms and therefore has a weight of two. For convenience we have also supplied the activation energy $\epsilon_{a}$ for each scattering process. Moreover, since the streamers in both electric fields are similar we only give the G-values corresponding to streamers in $12.5$\,kV/cm when they have reached a length of $6$\,mm.

\begin{table*}
    \centering
    \begin{tabular}{r||c|c}
         & G (1/(100 eV)) & $\epsilon_{a}$ (eV) \\
        \hline
N$_2(A^3\Sigma _u^+)$		&	1.055	&	6.169	\\
N$_2(B'^3\Sigma _u^-)$		&	0.104	&	8.166	\\
N$_2(B^3\Pi _g) $		&	1.007	&	7.354	\\
N$_2(C^3\Pi _u) $		&	0.741	&	11.033	\\
N$_2(E^3\Sigma _g^+)$		&	0.019	&	11.872	\\
N$_2(F^3\Pi _u) $		&	0.013	&	12.986	\\
N$_2(G^3\Pi _u) $		&	0.028	&	12.811	\\
N$_2(W^3\Delta_u) $		&	0.365	&	7.363	\\
N$_2(a''^1\Sigma _g^+)$		&	0.059	&	12.256	\\
N$_2(a'^1\Sigma _u^-)$		&	0.065	&	8.399	\\
N$_2(a^1\Pi _g) $		&	0.371	&	8.550	\\
N$_2(b'^1\Sigma _u^+)$		&	0.065	&	12.855	\\
N$_2(b^1\Pi _u) $		&	0.140	&	12.501	\\
N$_2(c_3^1\Pi _u) $		&	0.055	&	12.913	\\
N$_2(c_4^1\Sigma _u^+)$		&	0.038	&	12.935	\\
N$_2(o_3^1\Pi _u) $		&	0.034	&	13.104	\\
N$_2(w^1\Delta_u) $		&	0.063	&	8.896	\\
N$_2(J=2) $		&	34.406	&	0.002	\\
N$_2(v=1) $		&	13.834	&	0.288	\\
N$_2(v=2) $		&	6.337	&	0.573	\\
N$_2(v=3) $		&	3.368	&	0.855	\\
N$_2(v=4) $		&	1.731	&	1.133	\\
N$_2(v=5) $		&	0.822	&	1.408	\\
N$_2(v=6) $		&	0.365	&	1.679	\\
N$_2(v=7) $		&	0.152	&	1.947	\\
N$_2(v=8) $		&	0.059	&	2.211	\\
N$_2(v=9) $		&	0.021	&	2.471	\\
N$_2(v=10) $		&	0.007	&	2.728	\\
N + N 		&	0.749	&	9.754	\\
N$_2^+$		&	0.260	&	15.582	\\
N$_2^+(A^2\Pi _u) $		&	0.182	&	16.700	\\
N$_2^+(B^2\Sigma _u^+)$		&	0.040	&	18.752	\\
N + N$^+ $ 		&	0.023	&	24.342	\\
N + N$^{++}$		&	$3.5\cdot10^{-6}$	&	69.505	\\
    \end{tabular}
    \caption{The G-values for the generation of excited states of  \N{2}{}. The excitation energies of the species are given as well.}
    \label{tab:G_N2}
\end{table*}

\begin{table*}
    \centering
    \begin{tabular}{r||c|c}
         & G (1/(100 eV)) & $\epsilon_{a}$ (eV) \\
        \hline
O$_2($2B$)$ 	&	0.003	&	15.001	\\
O$_2(A^3\Sigma_u^+,\ A'^3\Delta_u,\ c^1\Delta_u)$	&	0.170	&	4.200	\\
O$_2(B^3\Sigma_u^-)$	&	0.594	&	6.120	\\
O$_2($LB$)$	&	0.022	&	15.001	\\
O$_2(a^1\Delta_g)$	&	0.276	&	0.977	\\
O$_2(b^1\Sigma_g^+)$	&	0.066	&	1.627	\\
O$_2(v=1$) 	&	51.701	&	0.214	\\
O$_2(v=2)$ 	&	7.296	&	0.460	\\
O$_2(v=3) $	&	0.150	&	0.696	\\
O + O 	&	0.341	&	5.100	\\
O$_2^-$	&	0.520	&	0.000	\\
O + O$^-$ 	&	0.012	&	0.000	\\
O$_2^+$ 	&	0.113	&	12.071	\\
O + O$^+$ 	&	0.013	&	23.002	\\
O + O$^{++}$	&	$1.3\cdot10^{-6}$	&	73.006	\\
    \end{tabular}
    \caption{The G-values for the generation of excited states of  \O{2}{}. The excitation energies of the species are given as well.}
    \label{tab:G_O2}
\end{table*}

\begin{table*}
    \centering
    \begin{tabular}{r||c|c}
         & G (1/(100 eV)) & $\epsilon_{a}$ (eV) \\
        \hline
CH$_4$($v=1,v=3$) 	&	2.299	&	0.362	\\
CH$_4$($v=2,v=4$) 	&	4.621	&	0.162	\\
CH$_3$ + H 	&	0.526	&	7.501	\\
CH$_2$ + H$_2$	&	0.205	&	9.101	\\
CH + H$_2$ + H 	&	0.003	&	15.501	\\
C + H$_2$ + H$_2$	&	$5.5\cdot10^{-4}$	&	15.501	\\
CH$_3$ + H$^-$ 	&	0.004	&	0.000	\\
H$_2$ + CH$_2^-$ 	&	$2.5\cdot10^{-4}$	&	0.000	\\

CH$_4^+$ 	&	0.144	&	12.631	\\
CH$_3$ + H$^+$ 	&	0.002	&	21.102	\\
H + CH$_3^+$ 	&	0.089	&	12.631	\\
CH$_2$ + H$_2^+$ 	&	$2.4\cdot10^{-4}$	&	22.302	\\
H$_2$ + CH$_2^+$ 	&	0.007	&	16.201	\\
H + H$_2$ + CH$^+$ 	&	0.002	&	22.202	\\
H$_2$ + H$_2$ + C$^+$ 	&	0.000	&	22.002	\\

    \end{tabular}
    \caption{The G-values for the generation of excited states of  \CH{4}{}. The excitation energies of the species are given as well.}
    \label{tab:G_CH4}
\end{table*}
\section*{References}
\bibliography{references, references_manual}
\end{document}